\documentclass[structabstract]{aa}  
\usepackage{graphicx}
\usepackage{txfonts}
\usepackage{longtable}

\usepackage{enumerate}
\usepackage{booktabs}

\begin{document}
   \title{Consequences of the simultaneous formation of giant planets by the core accretion mechanism}

   \subtitle{}

   \author{O. M. Guilera
          \inst{1,2}\fnmsep\thanks{Fellow of the Consejo Nacional de
	  Investigaciones  Cient\'{\i}ficas y T\'ecnicas (CONICET), Argentina.
          E-mail: oguilera@fcaglp.unlp.edu.ar}
          \and
          A. Brunini\inst{1,2}\fnmsep\thanks{Member of the Carrera del Investigador 
	  Cient\'{\i}fico, CONICET, Argentina. 
	  E-mail: abrunini@fcaglp.unlp.edu.ar}
          \and
	  O. G. Benvenuto
	  \inst{1,2}\fnmsep\thanks{Member of the Carrera del 
	  Investigador Cient\'{\i}fico, Comisi\'on de Investigaciones 
	  Cient\'{\i}ficas de la Provincia de Buenos Aires, Argentina.
	  E-mail: obenvenu@fcaglp.unlp.edu.ar}
          }
   \offprints{O. M. Guilera}

   \institute{Facultad de Ciencias Astron\'omicas y Geof\'{\i}sicas, 
              Universidad Nacional de La Plata, Paseo del Bosque s/n (B1900FWA) La Plata, 
              Argentina
              \and
	      Instituto de Astrof\'{\i}sica de La Plata, IALP, CCT-CONICET-UNLP, 
	      Argentina\\
              }

   \date{Accepted}

  \abstract
   {The core accretion mechanism is presently the most widely accepted cause of the formation of giant planets. For simplicity, most models presently assume that the growth of planetary embryos occurs in isolation.}
   {We explore how the simultaneous growth of two embryos at the present locations of Jupiter and Saturn affects the outcome of planetary formation.}
   {We model planet formation on the basis of the core accretion scenario and include several key physical ingredients. We consider a protoplanetary gas disk that exponentially decays with time. For planetesimals, we allow for a distribution of sizes from 100~m to 100~km with most of the mass in the smaller objects. We include planetesimal migration as well as different profiles for the surface density $\Sigma$ of the disk. The core growth is computed in the framework of the oligarchic growth regime and includes the viscous enhancement of the planetesimal capture cross-section. Planet migration is ignored.}
   {By comparing calculations assuming formation of embryos in isolation to calculations with simultaneous embryo growth, we find that the growth of one embryo generally significantly affects the other. This occurs in spite of the feeding zones of each planet never overlapping. The results may be classified as a function of the gas surface density profile $\Sigma$: if $\Sigma \propto r^{-3/2}$ and the protoplanetary disk is rather massive, Jupiter's formation inhibits the growth of Saturn. If $\Sigma \propto r^{-1}$ isolated and simultaneous formation lead to very similar outcomes; in the the case of $\Sigma \propto r^{-1/2}$ Saturn grows faster and induces a density wave that later acclerates the formation of Jupiter.}
   {Our results indicate that the simultaneous growth of several embryos impacts the final outcome and should be taken into account by planet formation models.}


   \maketitle
%

\section{Introduction}

\label{sec:intro}

The core instability mechanism is usually considered the way in which giant planets formation proceeds. This mechanism considers planetary formation 
as a consequence of a two-step process. A solid material seed with a mass of
the order of that of the Moon is immersed in a protoplanetary disk. This object
will be the core of the planet. The protoplanetary disk has a population of
planetesimals coexisting with a gaseous component. The Moon-sized object begins to accrete solid material from its feeding zone. The feeding zone is often assumed to be an annulus extended across few times the Hill radius $R_H$
(defined to be $R_H= a ( M / 3 M_* )^{1/3}$, where $a$ is the radius of the orbit
of the planet of mass $M$, while $M_*$ is the mass of the central star of the
system) on both sides of the orbit of the protoplanet. The core gravitationally
binds a tiny amount of gas and the whole protoplanet remains in hydrostatic
and thermal equilibrium. The gas accretion rate is iniatally far lower than
that of solids. This is true   until the core reaches a mass of $\approx 10
M_{\oplus}$, where $M_{\oplus}$ is the Earth mass. At these stages, the
amount of gas bound to the protoplanet becomes comparable to the core mass. It is then no longer possible to be in thermal equilibrium conditions and the protoplanetary envelope begings to contract. This corresponds to the onset of the runaway gas accretion. On a very short timescale, the planet accretes a large amount of gas
and reaches its final mass. This sequence of events was envisaged by Mizuno~(\cite{mizuno}) by employing static models and later with evolutionary models by Bodenheimer \& Pollack~(\cite{bodenheimer}) and Pollack et al.~(\cite{pollack}). Core instability calculations of giant planet formation have been carried out by many groups, e.g., Alibert et al.~(2005 \cite{alibert a}, \cite{alibert b}), Hubickyj et al.~(\cite{hubickyj}), and Dodson-Robinson et al.~(\cite{dr}).

Fortier et al.~(\cite{fbb}; \cite{fbb2}) were the first to consider the oligarchic growth regime for the accretion of planetesimals. This regime is known to provide a good description of planetesimal dynamics when the embryos present in the
nebula have masses much lower than the Earth mass. Thus, this is the correct
description of the accretion regime for almost the entire formation process.
Oligarchic growth takes into account the effects of the gravitational
perturbations of growing planets. In this regime, it is known that protoplanets
grow on a longer timescale that in other accretion regimes such as those
considered by, e.g., Pollack et al.~(\cite{pollack}). One of the most remarkable
characteristics of the core instability mechanism is that runaway gas accretion
occurs at a core mass value largely independent of the final mass of the
planet. This is in close agreement with the estimate of the core masses of the
giant planets of our Solar System (see Guillot, \cite{guillot} and Militzer et al., \cite{militzer}).

In spite of being considered the most likely mechanism of planetary
formation, core instability may have some serious difficulties. One of the main drawbacks was considered to be the long timescale for planetary formation. In many simulations, planets form on a time interval longer than (or need timescales uncomfortably close to) the dissipation timescale of the protoplanetary nebula, usually considered to be of the order of 10~Myr~(Hillenbrand, \cite{hillenbrand}). In any case, we should recall the number of simplifying assumptions
considered when constructing the models that lead to this apparent paradox.
For example, most of the  calculations available in the literature considered a
single size distribution of planetesimals. Relaxing this assumption
in favor of a size distribution for which most of the mass belongs to small
planetesimals, Benvenuto et al.~(\cite{bfb}) demonstrated that the four giant planets of the Solar System can be formed in a timely way with
core masses in close agreement with current estimate.

Another usual assumption in detailed simulations of planetary growth is that
each planet grows alone in the disk. This would be correct if the population of
planetesimals to be accreted by one planet were not appreciably perturbed by
the presence of another embryo. At first sight, it may be understood that this is the case if the feeding zone of each planet does not overlap the one corresponding to any other planet. However, this is {\it not} the case if
we include planetesimal migration. This process leads to a net inward motion
of planetesimals. A planet will perturb the swarm of planetesimals that may be
later accreted by another planet moving along an inner orbit. Moreover, as we show below, even the presence of an inner planet will be able to
affect the accretion process of an outer object. To our knowledge, no
calculation has been performed to quantitatively analyze these effects. It is
the main aim of the present paper to perform a first step towards filling this
gap. In doing so, we consider the simultaneous growth of two planets at
distances from the central star as those of Jupiter and Saturn. We consider a distribution of sizes of planetesimals and also different profiles
for the surface density of solids and gas corresponding to the protoplanetary
disk. We also consider two values of the gas to solid ratio.

Oligarchic growth predicts the simultaneous formation of many embryos on orbits
separated by about 10~$R_H$ from each other. Thus, starting with this initial
configuration would be more realistic than the one adopted here. However, to
perform a fully detailed simulation in this context we would have to be able to
compute the merging of embryos, which in the presence of a gaseous envelope for
each protoplanet is a very complex process. This is beyond the scope of the
present paper. In any case, even with the adopted initial configuration we
expect to realistically estimate the effect of the simultaneous growth
of more than one planet. In this work, we do not consider planetary migration. 

This paper is organized as follows: in Sect.~\ref{sec:diskmodel}, we describe the protoplanetary disk inside which giant planets grow. Section~\ref{sec:Oligarchic} is devoted to describing the oligarchic growth regime. Section~\ref{sec:mod_fom_planets} describes how we compute the growth of each protoplanetary embryo. With all the ingredients of our model quantitatively defined, in
Sect.~\ref{sec:simultaneous} we present our results corresponding to the formation of Jupiter and Saturn, comparing it with the cases of isolated formation. In Sect.~\ref{sec:exploring}, we explore different profiles for the surface density of the nebula, and in Sect.~\ref{sec:Hayahshi_Nebula} we apply our model to the standard Hayashi nebula. Finally, in Sect.~\ref{sec:disc} we present a discussion of our results and some concluding remarks.


\section{A brief description of the protoplanetary disk model} 
\label{sec:diskmodel}

We consider an axisymmetric disk with inner and outer radii of
$R_{min}=0.4$~AU and $R_{max}=30$~AU, respectively. The disk has both gaseous and
planetesimal components. For the surface density profile of both of them, we
consider a power law distribution of the form
\begin{eqnarray}
\Sigma_g &\propto& a^{-p} \\
\Sigma_s &\propto& a^{-p},
\end{eqnarray}
with a temperature profile given by
\begin{eqnarray}
T \propto a^{-1/2}.
\end{eqnarray}
It can be shown that the volumetric density distribution at the disk plane is
\begin{eqnarray}
\rho_g \propto a^{-p - 5/4}. 
\end{eqnarray}
Here we consider a density distribution for the size of planetesimals. We study 31 different sizes with radii between 100~m and 100~km in steps selected such that the quotient of masses of consecutive sizes is a factor of two, in a similar way to Brunini \& Benvenuto~(\cite{bb}). We assume that the material that composes the planetesimals has a density of $\rho_p= 1.5~g~cm^{-3}$. 

We consider a number of planetesimals per unit of mass distribution given by $dN/dm \propto m^{-5/2}$ (Kokubo \& Ida, \cite{kokubo1}, \cite{ki2000}; Ormel et al., \cite{ormel}), for which most of the mass of solids is in the smallest planetesimals ($M_T \propto m^{-1/2}$). Numerical simulations indicate that the mass distribution of planetesimals may be represented by a single or piecewise power law $dN/dm \propto m^{-q}$. For a constant value of $q$, the total mass of solids is $M_T \propto m^{2-q}$, so if $q<2$, the total mass of solids is contained in the biggest planetesimals; in contrast, if $q>2$ the total mass of solids is contained in the smallest planetesimals. Wetherill \& Stewart (\cite{ws}) studied the evolution of a planetesimal system considering an initial population of planetesimals with radius of $\sim$~10~km that evolved only by means of collisions and fragmentation. They found that a planetesimal size distribution relaxes to a piecewise power law: a population of small planetesimals produced by fragmentation (with $q\sim1.7$) and a population of large planetesimals that follow an accretive regime (with $q\sim2.5$). Kokubo \& Ida (\cite{ki2000}) studied the evolution of planetesimal size using N-body simulations finding that, in the oligarchic regime, large planetesimals follow a continuous power-law distribution with $2<q<3$. Ormel et al. (\cite{ormel}) performed statistical simulations that include several physical processes such as dynamical friction, viscous stirring, gas drag, and fragmentation, finding that the transition between the runaway growth and oligarchic growth is characterized by a power-law size distribution of mass index $q\sim2.5$. On the other hand, planetesimal formation continues to be studied and to date it is not well understood how meter to kilometer-sized planetesimals appear in the disk. Hence, the primordial size distribution of these bodies has not yet been established. In the paper of Wetherill \& Stewart (\cite{ws}), where the system evolves only through collisions and fragmentation, meter-sized planetesimals can only appear in the disk as fragments of larger bodies. Ida et al. (2008) performed some calculations showing that the magneto~-~rotational instability (Balbus \& Hawley \cite{balbus}; hereafter MRI) turbulence is a serious problem for planetesimal formation. They show that it is very difficult to form kilometer-sized planetesimals from centimeter-meter sized particles because of the predominance of an erosive rather than accretive regime. Under these conditions, it does not seem possible that kilometer-sized planetesimals formed by the accretion of smaller ones. However, they also suggest that, if MRI were inefficient, planetesimals may have formed in ``dead zones'', where the interaction between solids and gas is weak (Gammie, \cite{gammie}; Sano et al., \cite{sano}). This would favor the survival of centimeter and meter-sized bodies. Kilometer-sized planetesimals may have formed by the accretion of smaller planetesimals, and the coexistence of meter and kilometer sized planetesimals may have been possible. This possibility is important for the formation of the giant planets in the solar system, since smaller planetesimals favor the rapid formation of the solid embryos. Moreover, if our giant planets formed in an environment where the MRI was effective they should have migrated towards the Sun becoming hot giant planets (Matsumura et al., \cite{matsumura}). On the other hand, laboratory experiments show that reaccumulation of fragmentation debris can lead to the formation of planetesimals (Teiser \& Wurm, 2009). Owing to the present status of the theory of planetesimal formation, especially with regard to the period when they grow from meter to kilometer size, we consider that the size and mass distribution of the planetesimals adopted here is a valid hypothesis. 

Planetesimals are affected by gaseous friction, which causes an inwards decay and can alter the solid distribution in the disk. As demonstrated by Thommes et al.~(\cite{thommes}) and Chambers~(\cite{chambers}), this effect has a strong influence on the timescale of accretion and on the final mass values reached by the planets at different positions in the protoplanetary disk. The mean orbital evolution of planetesimals is given by (Adachi et al., \cite{adachi})
\begin{eqnarray}
\frac{da}{dt} = &-& \frac{2a}{T_{fric}}\bigg(\eta^2 + \frac{5}{8}e^2 +
 \frac{1}{2} i^2\bigg)^{1/2} \bigg[\eta + \bigg(\frac{5}{16} + \frac{\alpha}{4}\bigg)e^2 +
\frac{1}{4}i^2\bigg],
\end{eqnarray}
where 
\begin{eqnarray}
T_{fric}= \frac{8\rho_p r_p}{3C_D\rho_{g}v_k}.
\end{eqnarray}
Here $\rho_p$ is the density of planetesimals, $r_p$ is the radius of the planetesimal, $\eta=(v_k - v_{gas})/v_k$, and $\alpha$ is the exponent of the power-law density of the gas in the disk mid plane ($\rho_g \propto a^{-\alpha}$). 

The evolution of planetesimal disks follows the equation of continuity
\begin{eqnarray}
\frac{\partial \Sigma_s}{\partial t} - \frac{1}{a}\frac{\partial}{\partial a}
 \bigg(a\frac{da}{dt}\Sigma_s\bigg) = F(a), \label{eq:continuity}
\end{eqnarray}
where $F(a)$ describes the sinks of disk material (accretion by the forming planets and solid sublimation across the ice line, as done by Brunini \& Benvenuto, 2008). For simplicity, we also assume that the gaseous component dissipates following an exponential decay of its density as
\begin{eqnarray}
\rho_g(a,t)= \rho_g(a,0)e^{-t/\tau}.
\end{eqnarray}
We set $\tau=6$~Myr as the characteristic timescale for the dissipation of the protoplanetary nebula (Haisch et al., \cite{haisch}).


\section{Accretion onto planets: oligarchic growth regime}
\label{sec:Oligarchic}

The process of the accretion of solids is described by the ``particle in a box'' approximation (Inaba et al., \cite{inaba1})
\begin{eqnarray}
\frac{dM_C}{dt}= \frac{2\pi \Sigma(a_P)R_H^2}{P}P_{coll},
\label{eq:part_box} \end{eqnarray}
where $M_{C}$ is the mass of the core, $\Sigma(a_p)$ is the surface density of solids at the location of the planet, $R_H$ is the Hill radius, $P$ is the orbital period, and $P_{coll}$ is the collision probability, which is a function of the core radius, the Hill radius of the planet, and the relative velocity of planetesimals $P_{coll}=P_{coll}(R_C,R_H, v_{rel})$.  For regimes of high $(\hat{e},\hat{i}>2)$, medium $(0.2<\hat{e},\hat{i}<2)$ and low velocities $(\hat{e},\hat{i}<0.2)$, $P_{coll}$ is given by
\begin{eqnarray}
P_{coll\_high}&=& \frac{(R_C+r_p)^2}{2\pi R_H^2}\bigg[I_F(\beta) +
 \frac{6R_H I_G(\beta)}{(R_C+r_p)^2\hat{e}^2}\bigg], \\
P_{coll\_med}&=& \frac{(R_C+r_p)^2}{4\pi R_H^2 \hat{i}}\bigg[17.3 +
 \frac{232R_H}{(R_C+r_p)}\bigg], \\
P_{coll\_low}&=& 11.3\bigg[\frac{R_C+r_p}{R_H}\bigg]^{1/2},
\end{eqnarray}
where $R_C$ is the radius of the core, $\hat{e}$ and $\hat{i}$ are the reduced eccentricity and inclination, defined by $\hat{e}= ea_P/R_H$ and $\hat{i}= ia_P/R_H$, where $\beta=\hat{i}/\hat{e}$, $e$ and $i$ are the quadratic mean values of the eccentricity and inclination of planetesimals at the feeding zone of the planet, and $I_F(\beta)$ and $I_G(\beta)$ are functions that can be expressed in terms of elliptic integrals that, in the interval $0 < \beta \le 1$ can also be approximated  by (Chambers, \cite{chambers})
\begin{eqnarray}
I_F(\beta)&=& \frac{1+0.95925 \beta + 0.77251 \beta^2}{\beta(0.13142 +
 0.12295\beta)}, \\
I_G(\beta)&=& \frac{1 + 0.39960\beta}{\beta(0.0369 + 0.048333\beta +
 0.006874 \beta^2)}.
\end{eqnarray}

In contrast to the works of Brunini \& Benvenuto~(\cite{bb}) and Chambers~(\cite{chambers}), we adopt the results of Inaba et al.~(\cite{inaba1}), who assume that $P_{coll}$ is given in the whole range of velocities for $(\hat{e},\hat{i})$ as
\begin{eqnarray}
P_{coll}= \mbox{min}\big[P_{coll\_med}, (P_{coll\_low}^{-2} + P_{coll\_high}^{-2})^{-1/2}\big].
\end{eqnarray}
We consider the drag force that planetesimals experience on entering the planetary envelope, which largely increases their capture cross-section. Inaba \& Ikoma~(\cite{inaba2}) found an approximate solution to the equations of motion, which allows a rapid estimation of the critical radius for capture $r_p$ as a function of the radius of the captured planetesimal, the density of the gaseous envelope $\rho$, and the enhanced radius $\tilde{R}_C$
\begin{equation}
 r_p= \frac{3 \rho(\tilde{R}_C) \tilde{R}_C}{2\rho_p} \left(
 \frac{v_{\infty}^2 + 2GM_P(\tilde{R}_C) / \tilde{R}_C}{v_{\infty}^2 +
 2GM_P(\tilde{R}_C) / R_H}\right),
\end{equation}
where $v_{\infty}$ is the relative velocity of the planet and planetesimal when the two are far apart and $M_P(\tilde{R}_C)$ is the total mass of the planet contained whitin $\tilde{R}_C$. Inaba \& Ikoma~(\cite{inaba2}) propose replacing $\tilde{R}_C$ for $R_C$ in the expressions of collision probability
\begin{eqnarray}
P_{coll\_high}&=& \frac{(\tilde{R}_C+r_p)^2}{2\pi R_H^2}\bigg[I_F(\beta) +
 \frac{6R_H I_G(\beta)}{(\tilde{R}_C+r_p)^2\hat{e}^2}\bigg], \\
P_{coll\_med}&=& \frac{(\tilde{R}_C+r_p)^2}{4\pi R_H^2 \hat{i}}\bigg[17.3 +
 \frac{232R_H}{(\tilde{R}_C+r_p)}\bigg], \\
P_{coll\_low}&=& 11.3\bigg[\frac{\tilde{R}_C+r_p}{R_H}\bigg]^{1/2},
\end{eqnarray}
and again 
\begin{eqnarray}
P_{coll}= \mbox{min}\big[P_{coll\_med}, (P_{coll\_low}^{-2} + P_{coll\_high}^{-2})^{-1/2}\big], 
\end{eqnarray}
which gives $P_{coll}=P_{coll}(\tilde{R}_C,R_H, v_{rel})$. Finally, we derived a generalized version of Eq.~(\ref{eq:part_box}) following Brunini \& Benvenuto~(\cite{bb})
\begin{eqnarray}
\frac{dM_C}{dt}= \int_{DS} dm \int_{FZ} && 2\pi \psi(a, R_H, a_P) ~ \times \nonumber \\
&& \frac{2\pi\Sigma(a,m)R_H^2}{P}~P_{coll}(a,m)~a~da, 
\label{eq:part_box_generalized} 
\end{eqnarray}
where $DS$ represents the integration over the distribution of sizes of planetesimals and $FZ$ indicates that the integration extends over the feeding zone. The functional form of $\psi$ is the same as used by Brunini \& Benvenuto~(\cite{bb}).

\subsection{Relative velocities out of equilibrium} 
\label{sec:rel_vel_noeq}

The relative velocity, $v_{rel}$, between a planetesimal and the protoplanet may be described by
\begin{eqnarray}
v_{rel}= \sqrt{\frac{5}{8}e^2 + \frac{1}{2}i^2} v_k,
\end{eqnarray}
where $v_k$ is the Keplerian velocity at the distance $a_P$. The relative velocity is governed by gravitational stirring caused by the protoplanets and damping caused by gas drag. This can be modeled following Ohtsuki et al.~(\cite{ohtsuki}) as
\begin{eqnarray}
\bigg(\frac{d\langle e^2 \rangle}{dt}\bigg)_{stirr} &=&
 \bigg(\frac{M_P}{3bM_{\star}P}\bigg) P_{VS}, \\
\bigg(\frac{d\langle i^2 \rangle}{dt}\bigg)_{stirr} &=&
 \bigg(\frac{M_P}{3bM_{\star}P}\bigg) Q_{VS},
\end{eqnarray}
where 
\begin{eqnarray}
P_{VS} &=& \bigg[\frac{73\hat{e}^2}{10\Lambda^2}\bigg] \ln \bigg(1 +
 \frac{10\Lambda^2}{\hat{e}^2}\bigg) 
+ \bigg[\frac{72I_{PVS}(\beta)}{\pi\hat{e}\hat{i} }\bigg] \ln (1+\Lambda^2 ), \\
Q_{VS} &=& \bigg[\frac{4\hat{i}^2 + 0.2 \hat{i} \hat{e}^3}{10\Lambda^2
 \hat{e}}\bigg] \ln (1 + 10\Lambda^2) + \nonumber \\
 && \bigg[\frac{72I_{QVS}(\beta)}{\pi\hat{e}\hat{i} }\bigg] \ln(1+\Lambda^2),
\end{eqnarray}
where $\Lambda^2= \hat{i}(\hat{i}^2 + \hat{e}^2)/12$, and $I_{PVS}(\beta)$ and $Q_{PVS}(\beta)$ are given in terms of elliptic integrals than can be approximated by (Chambers, \cite{chambers})
\begin{eqnarray}
I_{PVS}(\beta) &=& \frac{\beta - 0.36251}{0.061547+0.16112\beta
 +0.054473\beta^2}, \\
Q_{PVS}(\beta) &=& \frac{0.71946-\beta}{0.21239+0.49764\beta
 +0.14369\beta^2}.
\end{eqnarray}
Nevertheless, these velocities decline appreciably when we move far away from the protoplanet. Hasegawa \& Nakazawa~(\cite{hn}) demonstrated that when the distance from the protoplanet is larger than $3.5 - 4$ times its Hill radius, the excitation of the relative velocities of planetesimals weakens significantly. Thus, we consider that
\begin{eqnarray}
 \bigg(\frac{d\langle e^2 \rangle}{dt}\bigg)_{stirr}^{efect} &=&
  f(\Delta)\bigg(\frac{d\langle e^2 \rangle}{dt}\bigg)_{stirr},
   \\
 \bigg(\frac{d\langle i^2 \rangle}{dt}\bigg)_{stirr}^{efect} &=&
  f(\Delta)\bigg(\frac{d\langle i^2 \rangle}{dt}\bigg)_{stirr}, 
\label{eq:stirring1}
\end{eqnarray}
where
\begin{eqnarray}
f(\Delta) = \Bigg(1 + \bigg|\bigg|
 \frac{\Delta}{4R_H}\bigg|\bigg|^5\Bigg)^{-1},
\label{eq:mod_fun}   
\end{eqnarray}
where $\Delta$ represents the distance from the protoplanet, and $f(\Delta)$ guarantees that the velocity profile of planetesimals is smooth along the entire disk and the planetary excitation on planetesimals is restricted to the protoplanetary neighborhood. This is important for an adequate solution of the solid distribution.

The friction caused by the gaseous component of the protoplanetary disk decreases the orbital eccentricities and inclinations of planetesimals at a rate given by (Adachi et al., \cite{adachi})
\begin{eqnarray}
\bigg(\frac{de}{dt}\bigg)_{gas} &=& \frac{\pi e
 r_p^2C_D\rho_{g}v_k}{2m_p}\bigg( \eta^2 + v_{rel}^2\bigg),  \\
\bigg(\frac{di}{dt}\bigg)_{gas} &=& \frac{\pi i
 r_p^2C_D\rho_{g}v_k}{4m_p}\bigg( \eta^2 + v_{rel}^2\bigg), 
\end{eqnarray}
where $C_D$ is a dimensionless coefficient that describes the gaseous friction (it is $\approx 1$ for the case of spherical bodies), $m_p$ is the mass of planetesimals, $\rho_{g}$ is the density of nebular gas at a distance $a$ from the central star, and $\eta$ is the ratio of the gas velocity to the local Keplerian velocity, given by
\begin{eqnarray}
\eta= \frac{v_k - v_{gas}}{v_k}= \frac{\pi}{16}(\alpha +
 \beta)\bigg(\frac{c_s}{v_k}\bigg)^2, 
\end{eqnarray}
where $\beta$ is the exponent of the power-law profile of temperature of the nebula ($T(a) \propto a^{-\beta}$) and $c_s$ is the local velocity of sound.


\section{Models of giant planet formation} \label{sec:mod_fom_planets}

As previously stated, the formation of planets was computed in the framework of the core instability scenario, in which three aspects are of key importance:
\begin{enumerate}[a.]
\item The solid accretion rate that increases the protoplanetary core mass. This is fundamental because it largely determines its formation timescale. The rate of solid accretion  also provides the energy release necessary to support the gaseous envelope against gravitational contraction. This rate was provided by the particle in a box approximation (Eq.~\ref{eq:part_box_generalized}) as described above.
\item The rate of gas accretion and the evolution of the envelope. The calculation of the structure of the gaseous protoplanetary envelope was performed by solving the standard equations of stellar structure (see below for additional details).
\item The interaction between planetesimals and the gaseous envelope. For simplicity, we assumed that they do not disaggregate on entering the protoplanetary envelope and are deposited in the nucleus. The gravitational energy release due to planetesimal accretion was incorporated at the bottom of the gaseous envelope.
\end{enumerate}

The equations that govern the evolution of the protoplanetary gaseous envelope are those of stellar evolution theory, namely
\begin{eqnarray}
\frac{\partial r}{\partial m_r} & = & \frac{1}{4\pi r^2
 \rho}\qquad \mbox{equation of definition of mass}
\\
\frac{\partial  P}{\partial  m_r} &  = &  -  \frac{G
m_r}{4\pi  r^4}\qquad \mbox{equation of hydrostatic equilibrium}
\\ 
\frac{\partial L_r}{\partial m_r} & = & \epsilon_{pl}-T\frac{\partial
 S}{\partial t}\qquad \mbox{equation of energetic balance} 
 \\
\frac{\partial T}{\partial m_r} &  = & - \frac{G m_r T}{4\pi r^4 P}
\nabla\qquad \mbox{equation of energy transport,}
\end{eqnarray}
where $\rho$ is the density of the envelope, $G$ is the universal gravitational constant, $\epsilon_{pl}$ is the energy release rate due to the accretion of planetesimals, S is the entropy per unit mass, and $\nabla \equiv \frac{d \ln T}{d \ln P}$ is the dimensionless temperature gradient, which depends on the type of energy transport.

We assumed that the outer boundary conditions change in full accordance with the evolution of the protoplanetary disk, i.e., are function of time. At the outermost point of the protoplanet model, we define $\rho= \rho_{neb}(t)$. However, we assumed the temperature to be fixed. The remainig ingredients of the calculations presented below are as in Fortier et al. (\cite{fbb}; \cite{fbb2}).


\subsection{Methodology of the code}
\label{sec:method_code}

The goal of this work was to construct self-consistent models of giant planet formation inside a realistic protoplanetary disk that evolves with time and, in particular, to analyze how the simultaneous growth of more than one object affects the growth of each individual object. For this purpose, we separated the calculation into several steps. We generalized the code that computes the in-situ formation of one giant planet to compute the simultaneous formation of N~planets. Then, we coupled this code with the one that solves the disk evolution, which, in turn, has the role of being the main program. 

Once the parameters model have been defined, the calculation starts with the initial disk model. The Hill's radius of each planet was then defined and the migration velocities computed for each planetesimal size. These velocities are employed to solve the surface density of the disk. Then we computed the evolution of the gaseous component of the disk. We then computed a growth step for each planet. To do so, we had to consider the contributions due to each planetesimal size inside the planetary feeding zone and also the collision probabilities for each planet. For each planet, we computed the accretion rate of solid and the new structure for the gaseous envelope. We adopted the new masses for each planet and computed the sinks of planetesimals, necessary for solving the continuity equation. This procedure took computational cycle, and when necessary the whole procedure was repeated. To achieve a stable and accurate sequence of models, for the next time step we adopted the minimum timescale of all the ingredients considered in the calculation.

We considered the simulation to have ended when all planets had reached the adopted mass values. In the case that one or more planets had reached their final masses while some others had not, we computed the evolution of the whole system. In this case, we did not consider any additional growth of the planets that reached their final masses but still took them into account when computing the evolution of the disk, perturbing the population of planetesimals.


\section{Application to the simultaneous formation of Jupiter and Saturn}
\label{sec:simultaneous}

We apply the model described in Sects. 2, 3 and 4 to quantitatively analyze the effects of the simultaneous formation of Jupiter and Saturn comparing with the results corresponding to the case of isolated formation.

\subsection{A standard nebula with profile $ \Sigma \propto r^{-3/2}$}

We first considered the standard model of the Solar nebula proposed by Hayashi~(\cite{hayashi}). It states that 
\begin{eqnarray}
\Sigma_s(a) &=& \left\lbrace
\begin{array}{ll}
7.1 \left(\displaystyle{\frac{a}{1~AU}}\right)^{-3/2}~g~cm^{-2}, & \mbox{$a<2.7$~AU}\\
\\
30  \left(\displaystyle{\frac{a}{1~AU}}\right)^{-3/2}~g~cm^{-2}, & \mbox{$a>2.7$~AU}
\end{array}
\right. \\
\Sigma_g(a) &=& 1700 \left(\frac{a}{1~AU} \right)^{-3/2}~g~cm^{-2} \\
T(a) &=& 280 \left(\frac{a}{1~AU}\right)^{-1/2}~K \\
\rho_g(a)&=& 1.4\times10^{-9} \left(\frac{a}{1~AU} \right)^{-11/4}~g~cm^{-3}
\end{eqnarray}

The discontinuity at $2.7$~AU in the surface densities of solids is caused by the condensation of volatiles (the ``snow line''). Hayashi~(\cite{hayashi}) employed a gas/solid ratio of 240 inside the snow line. Nevertheless, we employ the same model but with a  gas/solid ratio of 100 as in Mordasini et al.~(\cite{mordasini}). In that work, Mordasini et al.~(\cite{mordasini}) adopted a value of $z= 0.0149$ (Lodders, \cite{lodders}) for the abundance of heavy elements in the Sun; but this abundance value is higher by a factor of
2 to 4 in the internal regions of the Solar disk because of a redistribution of solids (Kornet et al., \cite{kornet}). Mordasini et al.~(\cite{mordasini}) adopted a factor of 3 considering a value of $z= 0.04$, which corresponds to a gas/sold ratio of 100 inside the snow line. Thus, our standard Solar nebula is defined by
\begin{eqnarray}
\Sigma_s(a) &=& \left\lbrace
\begin{array}{ll}
7.1 \left(\displaystyle{\frac{a}{1~AU}}\right)^{-3/2}~g~cm^{-2}, & \mbox{$a<2.7$~AU}\\
 \label{eq:dens_solid} \\
30  \left(\displaystyle{\frac{a}{1~AU}}\right)^{-3/2}~g~cm^{-2}, & \mbox{$a>2.7$~AU}
\end{array}
\right. \\
\Sigma_g(a) &=& 710 \left(\frac{a}{1~AU} \right)^{-3/2}~g~cm^{-2} \\
T(a) &=& 280 \left(\frac{a}{1~AU}\right)^{-1/2}~K \\
\rho_g(a)&=& 5.92 \times10^{-10} \left(\frac{a}{1~AU} \right)^{-11/4}~g~cm^{-3}.
\end{eqnarray}

For numerical reasons, we followed Thommes et al.~(\cite{thommes}) spreading the snow line in a region of about 1 AU with a smooth function, so the surface densities of solids are described by
\begin{eqnarray}
\Sigma_s(a) =  \left\lbrace 7.1 + \left( 30-7.1\right) \left[\frac{1}{2}\tanh\left(\frac{a-2.7}{0.5}\right) +  \frac{1}{2}\right]\right\rbrace && \times \nonumber \\
\left( \frac{a}{1 AU}\right) ^{-3/2}~g~cm^{-2} &&
\end{eqnarray}
This model evidently, corresponds to a nebula less massive than that of Hayashi~(\cite{hayashi}). In our model, the mass contained between 0.4 and 30~AU
\begin{eqnarray}
 0.005M_{\bigodot}\leq M_{gas} \leq 0.05M_{\bigodot},
\end{eqnarray}
which corresponds to nebulae of between 1 and 10~MMSN (where we assume that the minimum mass solar nebula corresponds to a disk of 0.005 solar masses).
\begin{table} 
\caption{Parameters for the planets considered in the calculations.}
\begin{center}
\begin{tabular}{lll}
\hline \hline & Jupiter & Saturn  \\\hline 
$R_{orb}$ [AU] & 5.2 & 9.5 \\
$M_{core}^{ini}$ [$M_{\oplus}$] & $5\times10^{-3}$ & $5\times10^{-3}$ \\
$M_{gas}^{ini}$ [$M_{\oplus}$] & $1\times10^{-12}$ & $1\times10^{-12}$ \\
$M_{final}$ [$ M_{\oplus}$] & 318 & 95 \\ \hline \hline
\end{tabular} 
\end{center}
\begin{list}{}{}
\item[$^{\mathrm{*}}$]{$R_{orb}$ is the radius of the (fixed) orbit. $M_{core}^{ini}$ and $M_{gas}^{ini}$ are the initial masses of the core and the envelope gaseous, respectively. $M_{final}$ is the final (assumed) mass value.}
\end{list}
\label{table:pla_ini_cond} 
\end{table}

\subsubsection{Jupiter and Saturn: isolated formation}
\label{sec:isolated}

After defining our disk model, we computed the isolated formation of Jupiter and Saturn. When we refer to isolated formation, we mean that we consider only one planet forms in the Solar nebula while the disk evolves. The initial conditions and final masses we considered here are given in Table~\ref{table:pla_ini_cond}.

The results we obtained are summarized in Table~\ref{table:IsoForSigma15}. We found that for 4 to 10~MMSN, Jupiter is formed in less than 6~Myr. Saturn is formed in less than 10~Myr for 6 to 10~MMSN. Although theoretical models find small cores for Jupiter (0-12~$M_{\oplus}$ Guillot, \cite{guillot}; 14-18~$M_{\oplus}$ Militzer et al., \cite{militzer}) and Saturn (9-22~$M_{\oplus}$ Guillot, \cite{guillot}), the models also predict $10-40~M_{\oplus}$ of heavy elements in Jupiter's envelope and $20-30 ~M_{\oplus}$ of heavy elements in Saturn's envelope (Guillot, \cite{guillot}; \cite{guillot2}). Because we assumed that all the falling planetesimals reach the core, $M_c$ corresponds to the total heavy element mass in the core and envelope. 
\begin{centering}
\begin{table}
\caption{Isolated formation of Jupiter and Saturn for a disk with surfaces densities of solids and gas $\propto a^{-3/2}$.}
\begin{tabular}{p{0.4cm}|p{0.58cm} p{0.58cm} p{0.58cm} p{0.65cm}|p{0.58cm} p{0.58cm} p{0.58cm} p{0.65cm}}
\toprule[0.8mm]

MM & & Jupiter & & & & Saturn & & \\
SN & & & & & &  & & \\ 

 \midrule[0.6mm]

 & $\Sigma_s$ 
 & $\rho_g$ 
 & $M_c$
 & $t_f$ 
 & $\Sigma_s$ 
 & $\rho_g$ 
 & $M_c$  
 & $t_f$ \\

\cmidrule[0.6mm](l){2-9}  
                 4 & 10.12 & 2.54 & 21.31 & 5.34 &  ---  & ---  &  ---  &  --- \\  
\midrule[0.3mm]  5 & 12.65 & 3.17 & 26.61 & 2.86 &  ---  & ---  &  ---  &  --- \\  
\midrule[0.3mm]  6 & 15.18 & 3.81 & 31.35 & 1.76 &  6.14 & 0.73 & 17.07 & 9.62 \\
\midrule[0.3mm]  7 & 17.71 & 4.45 & 36.32 & 1.09 &  7.17 & 0.85 & 20.83 & 4.89 \\
\midrule[0.3mm]  8 & 20.24 & 5.08 & 41.48 & 0.62 &  8.19 & 0.97 & 23.76 & 2.85 \\
\midrule[0.3mm]  9 & 22.77 & 5.72 & 45.39 & 0.33 &  9.22 & 1.09 & 26.38 & 1.75 \\
\midrule[0.3mm] 10 & 25.30 & 6.35 & 46.09 & 0.16 & 10.24 & 1.21 & 28.72 & 1.08 \\

\bottomrule[0.8mm] 
\end{tabular}
\begin{list}{}{}
\item[$^{\mathrm{*}}$]{$\Sigma_s~[g~cm^{-2}]$ and $\rho_g~[10^{-11}~g~cm^{-3}]$ are the initial surface density of solids and the initial volumetric density of gas at the disk mid-plane at the position of Jupiter and Saturn, respectively. $M_c~[M_{\oplus}]$ is the final core mass and $t_f~[Myr]$ is the formation time.}
\end{list}
\label{table:IsoForSigma15} 
\end{table} 
\end{centering}

\begin{centering}
\begin{table} 
\caption{Same as Table \ref{table:IsoForSigma15} but for the simultaneous formation of Jupiter and Saturn.} 
\begin{tabular}{p{0.4cm}|p{0.58cm} p{0.58cm} p{0.58cm} p{0.65cm}|p{0.58cm} p{0.58cm} p{0.58cm} p{0.65cm}}
\toprule[0.8mm]
MM & & Jupiter& & & & Saturn & & \\ 
SN & & & & & &  & & \\ 
\midrule[0.6mm]

 & $\Sigma_s$ & $\rho_g$ & $M_c$ & $t_f$ & $\Sigma_s$ & $\rho_g$ & $M_c$ & $t_f$ \\
\cmidrule[0.6mm](l){2-9} 
                 6 & 15.18 & 3.81 & 30.20 & 1.74 &  6.14 & 0.73 &  0.76 & $>10$ \\
\midrule[0.3mm]  7 & 17.71 & 4.45 & 34.67 & 1.07 &  7.17 & 0.85 &  2.91 & $>10$ \\
\midrule[0.3mm]  8 & 20.24 & 5.08 & 40.95 & 0.61 &  8.19 & 0.97 & 17.48 & 7.08 \\
\midrule[0.3mm]  9 & 22.77 & 5.72 & 44.48 & 0.33 &  9.22 & 1.09  & 22.26 & 3.02 \\
\midrule[0.3mm] 10 & 25.30 & 6.35 & 48.83 & 0.16 & 10.24 & 1.21  & 27.41 & 1.56 \\

\bottomrule[0.8mm] 
\end{tabular} 
\label{table:SimForSigma15} 
\end{table}
\end{centering}

\subsubsection{Jupiter and Saturn: simultaneous formation}
\label{sec:simultaneous2}

We next assumed that the planets form simultaneously; in this case, the main results we obtained are given in Table~\ref{table:SimForSigma15}. For 6 and 7~MMSN, we see that Jupiter formation completely inhibits Saturn formation (simulations are halted when the time of formation exceeds 10 Myr). The Saturn formation timescale is about 10 times longer than the Jupiter formation timescale for the remaining cases.

\begin{figure} 
\centering 
\includegraphics[height=0.4\textheight]{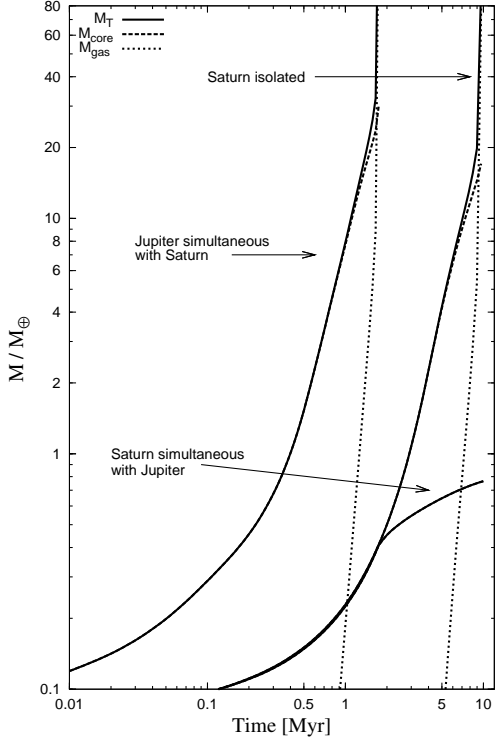} 
\caption{Comparison of cumulative masses as a function of time for the simultaneous formation of Jupiter and Saturn and isolated formation of Saturn for a 6 MMSN disk with power index $p= 3/2$. Clearly, when Jupiter reaches its gaseous run away, Saturn formation is inhibited in comparison with the case of its isolated formation.}
\label{fig:CompMasas6NM} 
\end{figure}

\begin{figure} 
\centering 
\includegraphics[height=0.4\textheight]{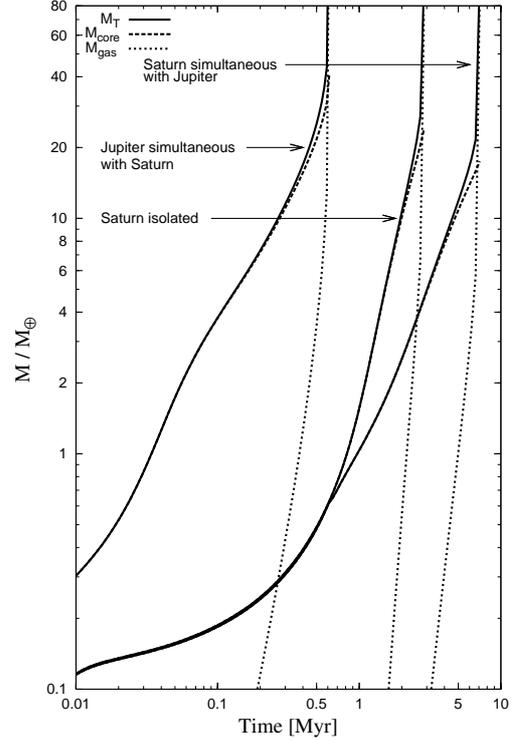} 
\caption{Same as Fig.~\ref{fig:CompMasas6NM} but for a 8 MMSN. In this case, Jupiter does not inhibit Saturn formation but largely delays it.}
\label{fig:CompMasas8NM} 
\end{figure}

\begin{figure} 
\centering 
\includegraphics[height=0.4\textheight]{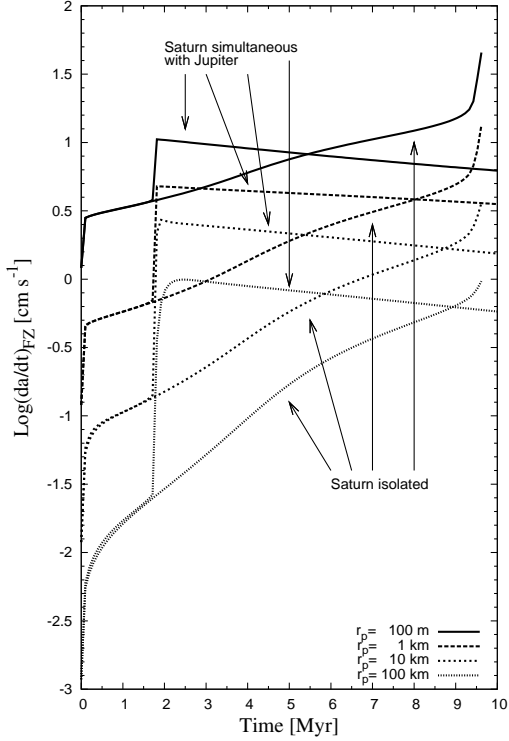} 
\caption{The evolution of planetesimal migration mean velocities at Saturn's feeding zone for a 6~MMSN disk with power index $p= 3/2$. The increase in the migration mean velocities, for the case of Saturn's simultaneous formation, is due to Jupiter's rapid formation as it reaches gaseous runaway.} 
\label{fig:CompVelos6NM} 
\end{figure}

\begin{figure} 
\centering 
\includegraphics[height=0.4\textheight]{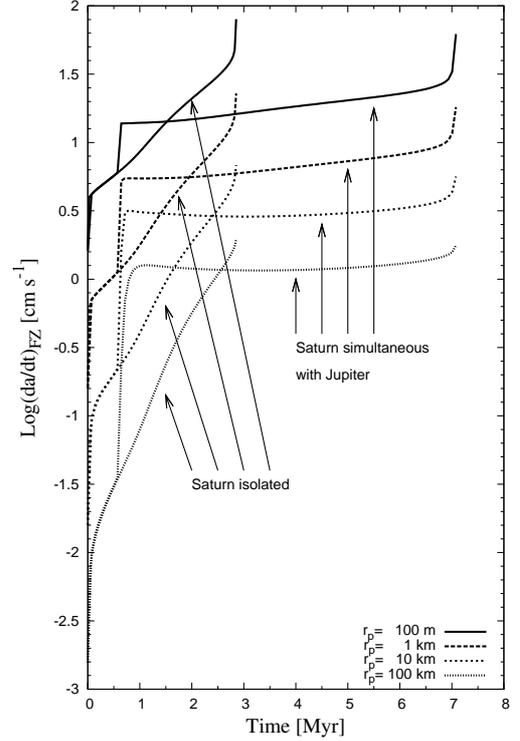} 
\caption{Same as Fig.~\ref{fig:CompVelos6NM} but for the case of a 8~MMSN nebula.}
\label{fig:CompVelos8NM}
\end{figure}

\begin{figure} 
\centering 
\includegraphics[height=0.4\textheight]{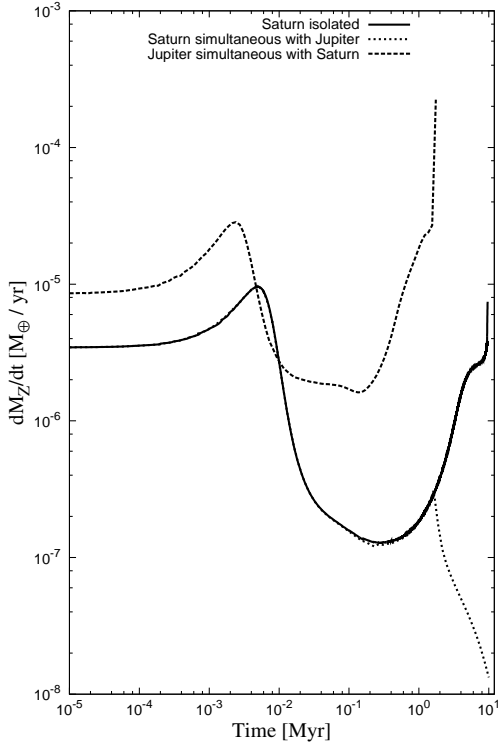} 
\caption{A comparison of the evolution of the solid accretion rate for the simultaneous formation of Jupiter and Saturn and isolated formation of Saturn for a 6~MMSN disk with power index $p= 3/2$. We see that Jupiter's formation cuts off the solid accretion rate in the case of simultaneous formation with Saturn.}
\label{fig:CompTasas6NM} 
\end{figure}

\begin{figure} 
\centering 
\includegraphics[height=0.4\textheight]{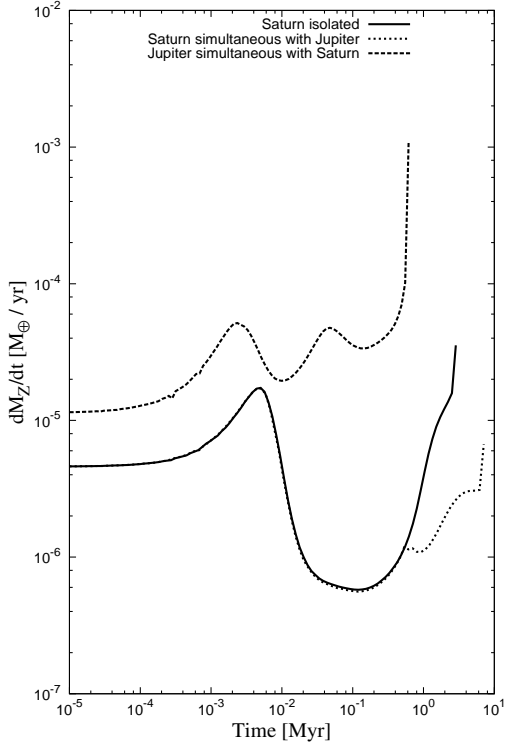} 
\caption{Same as Fig.~\ref{fig:CompTasas6NM} but for a 8 MMSN disk. The formation of Jupiter significantly modifies the solid accretion rate of Saturn, in constrast to the isolated formation, which delays Saturn's.} 
\label{fig:CompTasas8NM} 
\end{figure}

\begin{figure} 
\centering 
\includegraphics[height=0.4\textheight]{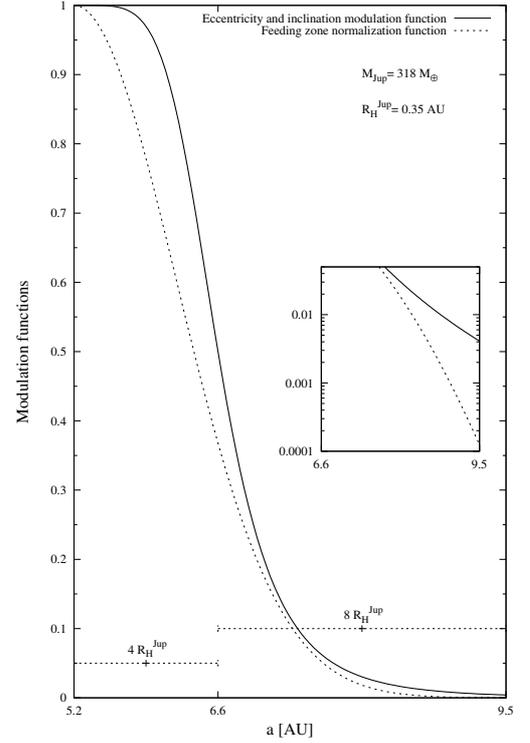}
\caption{Planetesimals's eccentricity and inclination modulation function due to Jupiter gravitational stirring (Eq.~\ref{eq:mod_fun}). When Jupiter achieves its final mass, the planetesimals's eccentricity and inclination excitation caused by Jupiter at Saturn's location, is less than 1~\% of its value at Jupiter's orbit.}
\label{fig:funcionmodulacion} 
\end{figure}

\begin{figure} 
\centering 
\includegraphics[height=0.5\textheight]{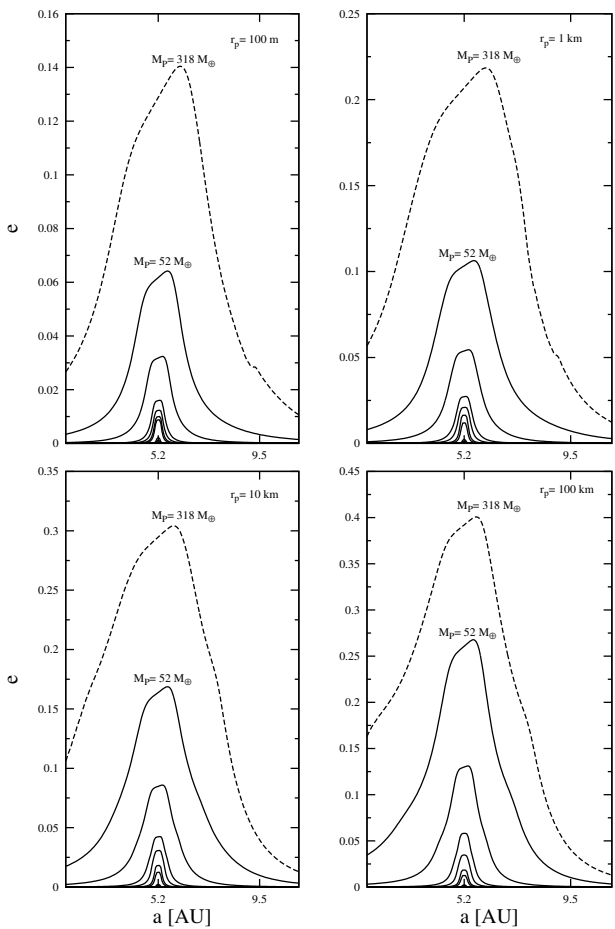}
\caption{Time evolution of the eccentricities along the disk for the case of the isolated formation of Jupiter. Curves correspond to $0, 1\times10^{-5}, 1\times10^{-4}, 1\times10^{-3}, 0.01, 0.05, 1, 1.73$, and $1.76$~Myr. The last two curves correspond to times before and after Jupiter gaseous runaway. At these moments, the mass of the planet was of $52~M_{\oplus}$ and $318~M_{\oplus}$, respectively.}
\label{fig:excentri} 
\end{figure}

\begin{figure} 
\centering 
\includegraphics[height=0.5\textheight]{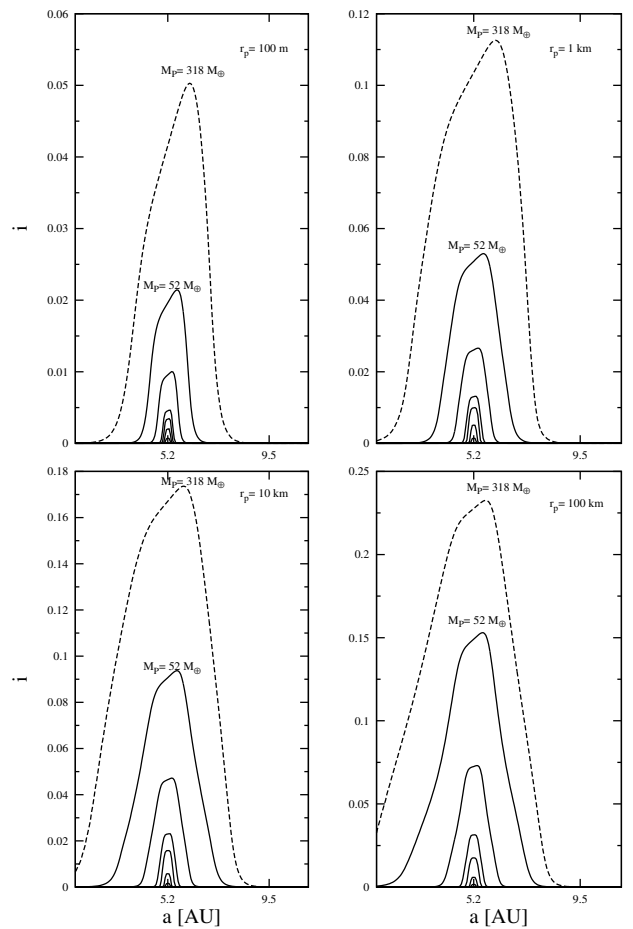}
\caption{Same as Fig.~\ref{fig:excentri}, but for the time evolution of the inclinations along the disk. It is clear that inclination excitations are smaller than eccentricity excitations.}
\label{fig:incli} 
\end{figure}

\begin{figure} 
\centering 
\includegraphics[height=0.5\textheight]{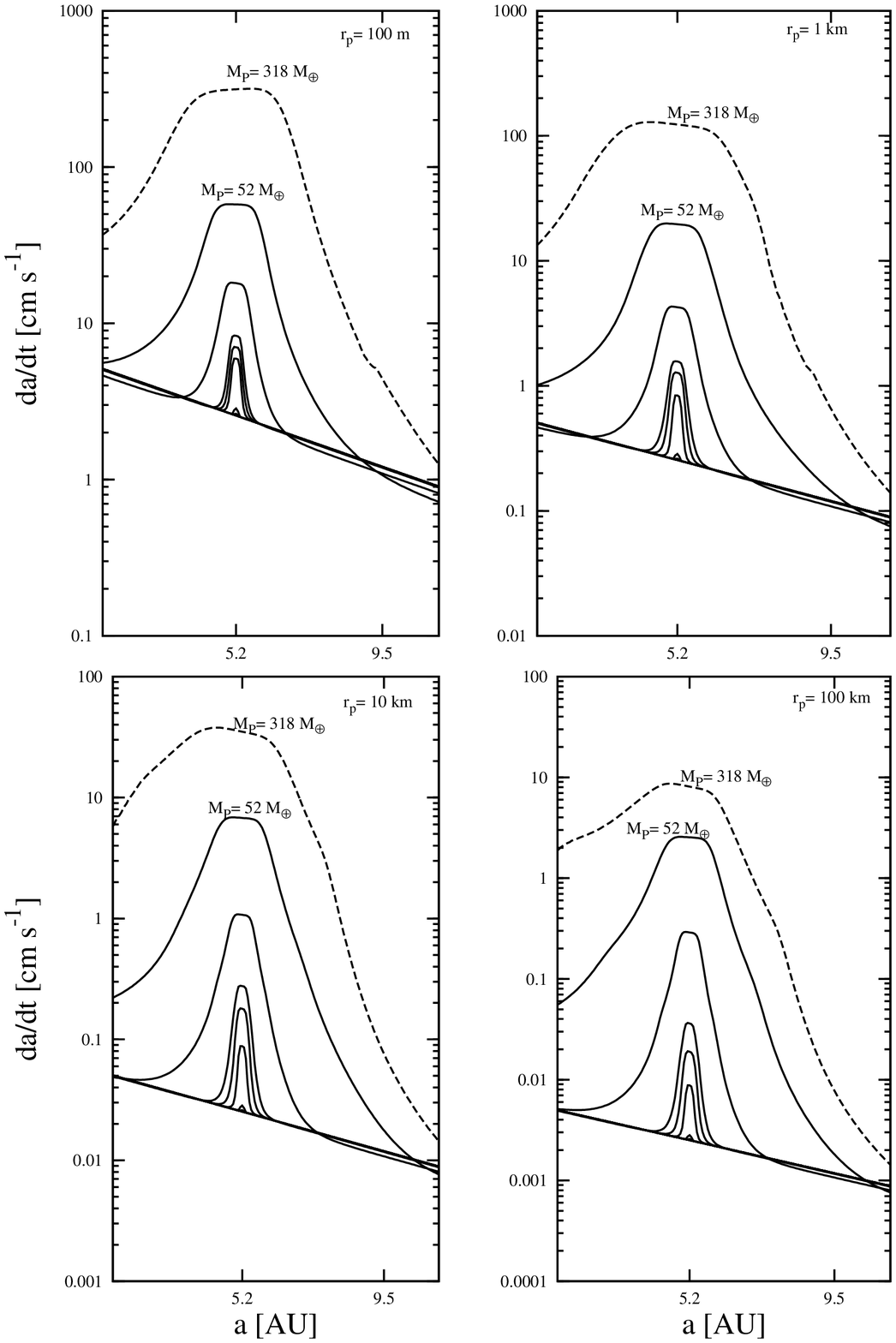}
\caption{Same as Figs.~\ref{fig:excentri} and \ref{fig:incli}, but for the time evolution of the planetesimal migration velocities along the disk. The planetesimal migration velocities, at 9.5~AU, are increased by Jupiter, only when it achieves its final mass.}
\label{fig:perfil_velo} 
\end{figure}

Comparing the results given in Tables~\ref{table:IsoForSigma15} and \ref{table:SimForSigma15}, we see that Saturn has almost no effect on the formation of Jupiter. However, the opposite is not true: the formation of Jupiter, clearly inhibits  the formation of Saturn in some cases (see Fig.~\ref{fig:CompMasas6NM}) and largely increases the formation time of Saturn in others (see Fig.~\ref{fig:CompMasas8NM}). This is due to an increment in the migration velocities of planetesimals in Saturn's neighborhood when Jupiter reaches gaseous runaway that, in turn, accelerates the migration of planetesimals to the Saturn's feeding zone (Figs.~\ref{fig:CompVelos6NM} and \ref{fig:CompVelos8NM}). This increment in the migration velocity of planetesimals causes the solid accretion timescale to become longer than planetesimal migration timescales, and the solid accretion rate of Saturn (when it is formed simultaneously with Jupiter) becomes less efficient than for the isolated Saturn formation (Figs.~\ref{fig:CompTasas6NM} and \ref{fig:CompTasas8NM}). Remarkably, Jupiter's and Saturn's feeding zones do not overlap at any time. Although Saturn's feeding zone is well beyond Jupiter's location, the increase in the migration velocity of planetesimals at this location, is caused by an eccentricity and inclination excitation of the planetesimals related to Jupiter's perturbations. When Jupiter reaches its final mass ($318~M_{\oplus}$), its corresponding Hill radius is $\sim 0.35$~AU. At 9.5~AU, Saturn is a distance of $\sim 12 R_H$ from Jupiter. As we can see in Fig.~\ref{fig:funcionmodulacion}, when Jupiter reaches its final mass, the tail of the modulation function (see Eq.~\ref{eq:stirring1}) produces an excitation (in planetesimal eccentricity and inclination) lower than 1\% at 9.5~AU compared to that produced at Jupiter's location. It is this excitation that causes the increment in the migration velocity of planetesimals at the Saturn's neighborhood when both planets are formed simultaneously. In Figs.~\ref{fig:excentri} and \ref{fig:incli}, we show the time evolution of eccentricity and inclination along the disk for the case of Jupiter formed in isolation. The increases in eccentricity and inclination when Jupiter reaches gaseous runaway, and reaches its final mass, are responsible for the increment in the  migration velocities of planetesimals at 9.5~AU (Fig.~\ref{fig:perfil_velo}). We note that the choice of the modulation function is rather arbitrary. However, it should satisfy some key properties: it should produce a planetary excitation on planetesimals restricted to the protoplanetary neighborhood; it {\it must} have a tail that is capable of exciting the surrounding planetesimals to be accreted; and, from a numerical point of view, it should guarantee a smooth velocity profile along the entire disk. Moreover, it ought to have an appreciable tail at distances well beyond the assumed function for representing the feeding zone ($\psi \propto e^{- (\frac{\Delta}{4R_H})^2}$). If this were not the case, there would be no planetary formation, even for the case of isolated embryos. The choice of another possible modulation function that satisfies the aforementioned conditions, but has a narrower tail would produce even smaller perturbations caused by Jupiter (in eccentricities and inclinations) at Saturn's location. However, if Jupiter and Saturn formed closer to each other, as predicted by the ``Nice Model'' of Tsiganis et al.~(\cite{tsiga}) (Jupiter$\sim$5.5 AU, Saturn$\sim$8.3 AU), an inhibition in the formation of Saturn should still occur. To our knowledge, a detailed study of this modulation function is not yet available and its calculation is a difficult task, beyond the scope of this paper. In any case, it seems an unavoidable conclusion that the modulation function should be significantly higher than zero on a long tail, sufficient to allow the excitation of planetesimals by an already formed Jupiter at Saturn's orbit. 

When the formation timescales for Jupiter are much shorter than those corresponding to Saturn, the formation of the latter is inhibited. One possible way out of this effect is to consider a smoother profile for the nebular surface density. This point is investigated in the next section.

\section{Exploring different disk profiles}
\label{sec:exploring}

Models of steady accretion $\alpha$-disks predict profiles of $\Sigma \propto r^{-1}$ (Shakura \& Sunyaev, \cite{ss}; Pringle, \cite{pringle}). We used this profile and the profile $\Sigma \propto r^{-1/2}$, previously proposed by Lissauer~(\cite{lissauer87}). We repeated the calculations reported in Sect.~\ref{sec:simultaneous}, but using instead these profiles for the disk density.

We decided to normalize these profiles at Jupiter's location. In this way, for the smoother slopes of the nebular surface densities, Saturn will find more material inside its corresponding feeding zone and, in turn, the formation timescales will be shorter. We note that in this case the mass of the disks between 0.4-30~AU never exceeds 0.1~$M_{\odot}$, so these disks should be stable against self-gravitational collapse (Mayer et al., \cite{mayer}).

We have shown that for our model of Solar nebula, the surface density is given by Eq.~\ref{eq:dens_solid}. It can be rewritten, scaling to Jupiter's location as
\begin{eqnarray}
\Sigma_s(a) &=& \left\lbrace
\begin{array}{ll}
0.6  \left(\displaystyle{\frac{a}{5.2~AU}}\right)^{-3/2}~g~cm^{-2}, & \mbox{$a<2.7$~AU}\\
\\
2.53 \left(\displaystyle{\frac{a}{5.2~AU}}\right)^{-3/2}~g~cm^{-2}, & \mbox{$a>2.7$~AU}.
\end{array}
\right. 
\end{eqnarray}
Thus, we employ a solid profile
\begin{eqnarray}
\Sigma_s(a) &=& \left\lbrace
\begin{array}{ll}
0.6  \left(\displaystyle{\frac{a}{5.2~AU}}\right)^{-p}~g~cm^{-2}, & \mbox{$a<2.7$~AU}\\
\\
2.53 \left(\displaystyle{\frac{a}{5.2~AU}}\right)^{-p}~g~cm^{-2}, & \mbox{$a>2.7$~AU}.
\end{array}
\right. \label{eq:dens_solid_norm}
\end{eqnarray}
Using the same profile for the temperature and the same gas/solid ratio for the different values of $p$ (-3/2, -1, -1/2) at Jupiter's location (5.2~AU), the value of the surface densities of solids and gas and the volumetric gas density in the mid plane of the disk were found to be the same regardless of the value of $p$.

 \subsection{A nebula with profile $\Sigma \propto r^{-1}$}

For example, we considered $p=1$ in the profile given by Eq. \ref{eq:dens_solid_norm}

\begin{eqnarray}
\Sigma_s(a) &=& \left\lbrace
\begin{array}{ll}
0.6  \left(\displaystyle{\frac{a}{5.2~AU}}\right)^{-1}~g~cm^{-2}, & \mbox{$a<2.7$~AU}\\
\\
2.53 \left(\displaystyle{\frac{a}{5.2~AU}}\right)^{-1}~g~cm^{-2}, & \mbox{$a>2.7$~AU}.
\end{array}
\right. 
\end{eqnarray}
Rescaling to 1~AU and using the same gas to solid ratio, we found that the disk becomes defined as
\begin{eqnarray}
\Sigma_s(a) &=& \left\lbrace
\begin{array}{ll}
3.13 \left(\displaystyle{\frac{a}{1~AU}}\right)^{-1}~g~cm^{-2}, & \mbox{$a<2.7$~AU}\\
\\
13.15  \left(\displaystyle{\frac{a}{1~AU}}\right)^{-1}~g~cm^{-2}, & \mbox{$a>2.7$~AU}
\end{array}
\right. \\
\Sigma_g(a) &=& 313 \left(\frac{a}{1~AU} \right)^{-1}~g~cm^{-2} \\
T(a) &=& 280 \left(\frac{a}{1~AU}\right)^{-1/2}~K \\
\rho_g(a)&=& 2.61 \times10^{-10} \left(\frac{a}{1~AU} \right)^{-9/4}~g~cm^{-3}.
\end{eqnarray}
Spreading the snow line in a region of about 1 AU with a smooth function
\begin{eqnarray}
\Sigma_s(a) = \left\lbrace 3.13 + \left(13.15-3.13\right) \left[ \frac{1}{2}\tanh\left(\frac{a-2.7}{0.5}\right) + \frac{1}{2}\right]\right\rbrace && \times  \nonumber \\
\left( \frac{a}{1 AU}\right)^{-1}~g~cm^{-2}, &&
\end{eqnarray}
the disk mass enclosed between 0.4 and 30~AU was found to be
\begin{eqnarray}
0.0065M_{\bigodot}\leq M_{gas} \leq 0.065M_{\bigodot},
\end{eqnarray}
which corresponds to values for nebulae between 1 and 10~MMSN (now the MMSM correspond to a disk of 0.0065 solar masses).

After defining the disk, we computed planetary formation.


\subsubsection{Jupiter and Saturn: isolated formation}

The results we obtained are presented in Table~\ref{table:IsoForSigma1}. For 3 to 10 MMSN, Jupiter's formation time was found to be short enough not to conflict with observational estimations. However, the final core mass of Jupiter is higher than expected. Similar results were obtained for the isolated formation of Saturn. We found that the timescales for the isolated formation of Jupiter and Saturn (except for the case of 4~MMSN) are in closer agreement (in all cases, Jupiter is formed before Saturn). We next considered the case of simultaneous formation.


\subsubsection{Jupiter and Saturn: simultaneous formation} 

Comparing Tables~\ref{table:IsoForSigma1} and \ref{table:SimForSigma1}, we see that, except for the case of 4~MMSN, for which the formation timescales are very different, in the remaining cases the simultaneous formation of Jupiter and Saturn is analogous to the case of isolated growth. 

Although in all cases, the final core masses are higher than observational estimations, we note that ablation of accreted planetesimals was not considered here. Small planetesimals could completely disintegrate before they reach the core, reducing the final core masses of Jupiter and Saturn.

\begin{centering}
\begin{table} 
\caption{Isolated formation of Jupiter and Saturn for a disk with power index $p= 1$.}
\begin{tabular}{p{0.4cm}|p{0.58cm} p{0.58cm} p{0.58cm} p{0.65cm}|p{0.58cm} p{0.58cm} p{0.58cm} p{0.65cm}}
\toprule[0.8mm]

MM & & Jupiter & & & & Saturn & & \\
SN & & & & & &  & & \\ 

 \midrule[0.6mm]

 & $\Sigma_s$ 
 & $\rho_g$ 
 & $M_c$
 & $t_f$ 
 & $\Sigma_s$ 
 & $\rho_g$ 
 & $M_c$  
 & $t_f$ \\

\cmidrule[0.6mm](l){2-9} 

                 3 &  7.59 & 1.90 & 21.20 & 6.87 &  ---  & ---  &   --- &  --- \\
\midrule[0.3mm]  4 & 10.12 & 2.54 & 26.87 & 3.34 &  5.53 & 0.66 & 19.41 & 8.55 \\
\midrule[0.3mm]  5 & 12.65 & 3.17 & 31.35 & 2.06 &  6.92 & 0.82 & 24.20 & 3.79 \\
\midrule[0.3mm]  6 & 15.18 & 3.81 & 34.90 & 1.37 &  8.30 & 0.99 & 28.02 & 2.00 \\
\midrule[0.3mm]  7 & 17.71 & 4.45 & 38.40 & 0.89 &  9.69 & 1.15 & 31.05 & 1.06 \\
\midrule[0.3mm]  8 & 20.24 & 5.08 & 42.01 & 0.51 & 11.07 & 1.31 & 33.87 & 0.56 \\
\midrule[0.3mm]  9 & 22.77 & 5.72 & 44.76 & 0.25 & 12.45 & 1.48 & 36.83 & 0.30 \\
\midrule[0.3mm] 10 & 25.30 & 6.35 & 50.20 & 0.11 & 13.84 & 1.64 & 41.10 & 0.18 \\

\bottomrule[0.8mm] 
\end{tabular} 
\label{table:IsoForSigma1}
\end{table}
\end{centering}

\begin{centering}
\begin{table} 
\caption{Same as Table~\ref{table:IsoForSigma1} but for the simultaneous formation of Jupiter and Saturn.}
\begin{tabular}{p{0.4cm}|p{0.58cm} p{0.58cm} p{0.58cm} p{0.65cm}|p{0.58cm} p{0.58cm} p{0.58cm} p{0.65cm}}
\toprule[0.8mm]

MM & & Jupiter & & & & Saturn & & \\
SN & & & & & &  & & \\ 

 \midrule[0.6mm]

 & $\Sigma_s$ 
 & $\rho_g$ 
 & $M_c$
 & $t_f$ 
 & $\Sigma_s$ 
 & $\rho_g$ 
 & $M_c$  
 & $t_f$ \\

\cmidrule[0.6mm](l){2-9} 

                 4 & 10.12 & 2.54 & 27.30 & 3.27 &  5.53 & 0.66 &  4.28 & $> 10$ \\
\midrule[0.3mm]  5 & 12.65 & 3.17 & 29.22 & 2.01 &  6.92 & 0.82 & 24.32 & 3.93 \\
\midrule[0.3mm]  6 & 15.18 & 3.81 & 33.73 & 1.33 &  8.30 & 0.99 & 29.67 & 1.96 \\
\midrule[0.3mm]  7 & 17.71 & 4.45 & 38.35 & 0.86 &  9.69 & 1.15 & 32.91 & 1.08 \\
\midrule[0.3mm]  8 & 20.24 & 5.08 & 45.52 & 0.49 & 11.07 & 1.31 & 34.51 & 0.56 \\
\midrule[0.3mm]  9 & 22.77 & 5.72 & 48.14 & 0.23 & 12.45 & 1.48 & 39.34 & 0.30 \\
\midrule[0.3mm] 10 & 25.30 & 6.35 & 49.95 & 0.10 & 13.84 & 1.64 & 44.54 & 0.19 \\
\bottomrule[0.8mm] 
\end{tabular}  
\label{table:SimForSigma1}
\end{table}
\end{centering}


\subsection{A nebula with profile $\Sigma \propto r^{-1/2}$}
 
For the profile given by Eq. \ref{eq:dens_solid_norm} with power of $p=1/2$, rescaling at 1~AU, employing the same gas/solid ratio, and spreading the snow line in a region of about 1 AU with a smooth function we found that the disk is characterized by
\begin{eqnarray}
\Sigma_s(a) &=& \left\lbrace 1.36 + \left(5.77-1.36\right) \left[ \frac{1}{2}\tanh\left(\frac{a-2.7}{0.5}\right) + \frac{1}{2}\right]\right\rbrace \times \nonumber \\ 
&& \left( \frac{a}{1 AU}\right) ^{-1/2} g/cm^{-2}, \\
\Sigma_g(a) &=& 136 \left(\frac{a}{1~AU} \right)^{-1/2}~g~cm^{-2}, \\
T(a) &=& 280 \left(\frac{a}{1~AU}\right)^{-1/2}~K, \\
\rho_g(a)&=& 1.13 \times10^{-10} \left(\frac{a}{1~AU} \right)^{-7/4}~g~cm^{-3}.
\end{eqnarray}
The mass of gas between 0.4~AU and 30~AU is
\begin{eqnarray}
 0.01M_{\bigodot}\leq M_{gas} \leq 0.1M_{\bigodot},
\end{eqnarray}
which correspond to nebulae of masses between 1 and 10~MMSN (where the MMSM correspond to a disk of 0.01 solar masses).

Again, we first compute isolated formation for Jupiter and Saturn.

\subsubsection{Jupiter and Saturn: isolated formation}

\begin{centering}
\begin{table} 
\caption{Isolated formation of Jupiter and Saturn for a disk with power index $p= 1/2$.}
\begin{tabular}{p{0.4cm}|p{0.58cm} p{0.58cm} p{0.58cm} p{0.65cm}|p{0.58cm} p{0.58cm} p{0.58cm} p{0.65cm}}
\toprule[0.8mm]

MM & & Jupiter & & & & Saturn & & \\
SN & & & & & &  & & \\ 

 \midrule[0.6mm]

 & $\Sigma_s$ 
 & $\rho_g$ 
 & $M_c$
 & $t_f$ 
 & $\Sigma_s$ 
 & $\rho_g$ 
 & $M_c$  
 & $t_f$ \\

\cmidrule[0.6mm](l){2-9}

                 3 &  7.59 & 1.90 & 26.41 & 4.80 &  5.61 & 0.66 & 23.92 & 5.92 \\
\midrule[0.3mm]  4 & 10.12 & 2.54 & 31.02 & 2.66 &  7.48 & 0.88 & 29.41 & 2.33 \\
\midrule[0.3mm]  5 & 12.65 & 3.17 & 34.49 & 1.75 &  9.36 & 1.09 & 33.48 & 0.99 \\
\midrule[0.3mm]  6 & 15.18 & 3.81 & 37.44 & 1.21 & 11.23 & 1.31 & 37.60 & 0.41 \\
\midrule[0.3mm]  7 & 17.71 & 4.45 & 39.82 & 0.79 & 13.10 & 1.53 & 43.02 & 0.19 \\
\midrule[0.3mm]  8 & 20.24 & 5.08 & 41.91 & 0.43 & 14.97 & 1.75 & 48.52 & 0.11 \\
\midrule[0.3mm]  9 & 22.77 & 5.72 & 44.40 & 0.19 & 16.24 & 1.97 & 53.34 & 0.07 \\
\midrule[0.3mm] 10 & 25.30 & 6.35 & 51.66 & 0.07 & 18.72 & 2.19 & 56.98 & 0.05 \\

\bottomrule[0.8mm] 
\end{tabular} 
\label{table:IsoForSigma05} 
\end{table} 
\end{centering}

\begin{centering}
\begin{table} 
\caption{Same as Table \ref{table:IsoForSigma05} but for the simultaneous formation of Jupiter and Saturn.}
\begin{tabular}{p{0.4cm}|p{0.58cm} p{0.58cm} p{0.58cm} p{0.65cm}|p{0.58cm} p{0.58cm} p{0.58cm} p{0.65cm}}
\toprule[0.8mm]

MM & & Jupiter & & & & Saturn & & \\
SN & & & & & &  & & \\ 

 \midrule[0.6mm]

 & $\Sigma_s$ 
 & $\rho_g$ 
 & $M_c$
 & $t_f$ 
 & $\Sigma_s$ 
 & $\rho_g$ 
 & $M_c$  
 & $t_f$ \\

\cmidrule[0.6mm](l){2-9} 

                 3 &  7.59 & 1.90 & 33.94 & 4.51 &  5.61 & 0.66 & 23.60 & 5.91 \\
\midrule[0.3mm]  4 & 10.12 & 2.54 & 43.92 & 2.30 &  7.48 & 0.88 & 28.42 & 2.33 \\
\midrule[0.3mm]  5 & 12.65 & 3.17 & 62.54 & 1.08 &  9.36 & 1.09 & 33.49 & 0.99 \\
\midrule[0.3mm]  6 & 15.18 & 3.81 & 60.07 & 0.50 & 11.23 & 1.31 & 37.58 & 0.41 \\
\midrule[0.3mm]  7 & 17.71 & 4.45 & 59.94 & 0.26 & 13.10 & 1.53 & 43.20 & 0.19 \\
\midrule[0.3mm]  8 & 20.24 & 5.08 & 57.53 & 0.16 & 14.97 & 1.75 & 48.75 & 0.11 \\
\midrule[0.3mm]  9 & 22.77 & 5.72 & 57.77 & 0.10 & 16.24 & 1.97 & 53.76 & 0.07 \\
\midrule[0.3mm] 10 & 25.30 & 6.35 & 57.90 & 0.06 & 18.72 & 2.19 & 57.60 & 0.05 \\
\bottomrule[0.8mm] 
\end{tabular}
\label{table:SimForSigma05} 
\end{table} 
\end{centering}

For each planet, formation timescales are again, shorter than estimate of disk lifetimes, but the final core masses are higher than observational estimate.

Except for the case of 3~MMSN, Saturn formation occurs remarkably on a timescale shorter than that corresponding to Jupiter. This result may seem rather paradoxical because the solid accretion rate is lower at larger distances from the central star~(Eq. \ref{eq:part_box}), even for this moderate profile. However, it has a simple explanation. For a solid density profile $\Sigma_s \propto r^{-p}$, if we consider an annulus of the disk of thickness $dr$, the mass of the annulus is $dm= 2\pi r\Sigma_s(r)dr$ so, $dm \propto r^{1-p}$. If $p>1$, the mass increases inwards and due to the planetesimal migration, the incoming flux of mass towards the annulus is lower than the outcoming one. On the other hand, if $p<1$, the mass grows outwards, and the incoming flux of mass is greater than the outcoming. The evolution of the solid density profile is governed by Eq.~\ref{eq:continuity}, where sinks, caused by planetary accretion, are also considered. Figure~\ref{fig:CompDensMedias6NMSigma05} clearly shows that initially the mean density of solids in Jupiter's feeding zone is greater than the mean density of solids in Saturn's feeding zone, but as time advances, the mean density of solids in Saturn's feeding zone become greater than that corresponding to Jupiter. Because of this, Saturn's formation occurs before that of Jupiter's. It is clear that the migration of planetesimals plays a very important role in planetary formation, especially for $p <1$.

We now consider simultaneous formation. 

\subsubsection{Jupiter and Saturn: simultaneous formation} 

\begin{figure} 
\centering 
\includegraphics[height=0.4\textheight]{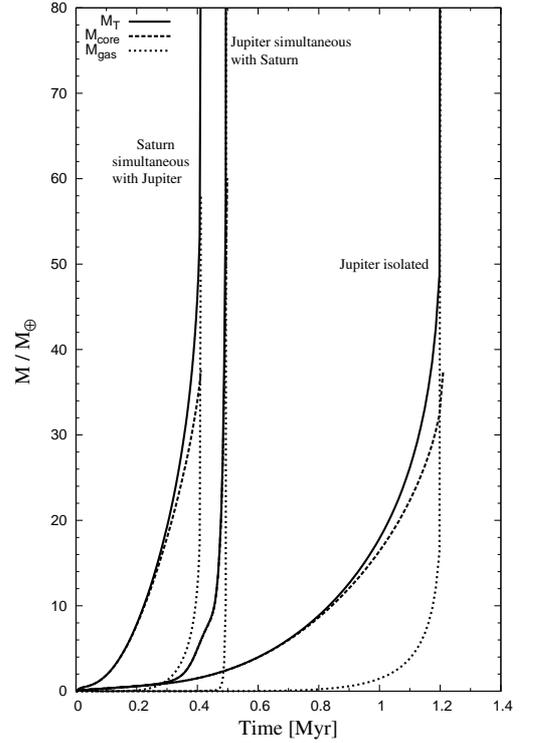} 
\caption{Comparison of cumulative masses as a function of time for the simultaneous formation of Jupiter and Saturn and isolated formation of Jupiter for a 6 MMSN disk with power index $p= 1/2$. The rapid formation of Saturn significantly favors the formation of Jupiter.}
\label{fig:CompMasas6NMSigma05}
\end{figure}

We see that for this nebular profile the temporal order of the phenomena is inverted. Jupiter formation does not affect the formation of Saturn (the latter forms first except in the case of 3~MMSN), but Saturn formation shortens the Jupiter formation timescale and largely increases their final cores (Fig.~\ref{fig:CompMasas6NMSigma05}).

\begin{figure} 
\centering 
\includegraphics[height=0.4\textheight]{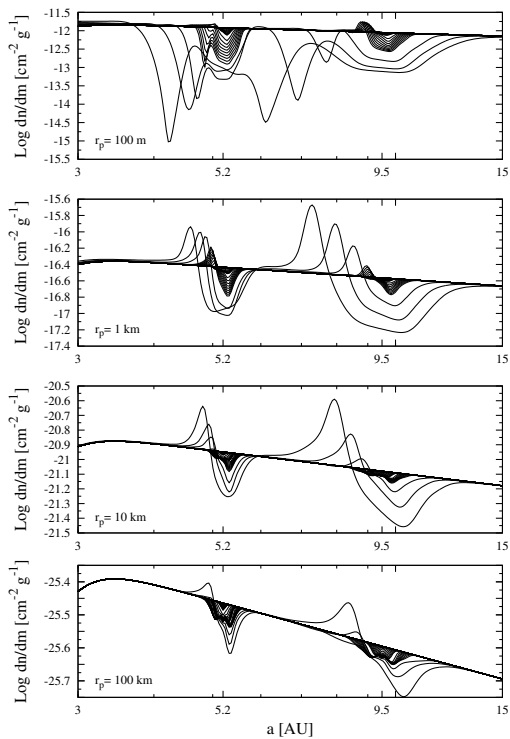} 
\caption{Time evolution of the planetesimal density profiles for a 6~MMSN disk with power index $p= 1/2$ for the simultaneous formation of Jupiter and Saturn. The density wave of solids, produced by the rapid formation of Saturn is a very dispersive phenomena and strongly depends on planetesimal size. As time advances, density profiles decrease at the planets's location. Curves correspond to $0, 1\times10^{-7}, 1\times10^{-6}, 1\times10^{-5}, 1\times10^{-4}, 1\times10^{-3}, 0.01, 0.05, 0.1, 0.2, 0.3, 0.4$, and $0.5$~Myr.}
\label{fig:CompPerfiles6NMSigma05_2} 
\end{figure}

\begin{figure} 
\centering 
\includegraphics[height=0.4\textheight]{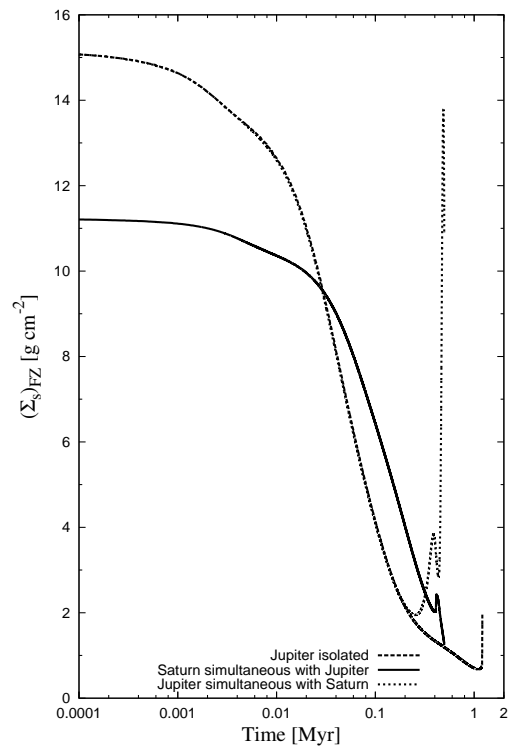} 
\caption{Evolution of the mean density of solids in the Jupiter and Saturn feeding zones for a 6~MMSN disk with power index $p= 1/2$. It is clear that the presence of Saturn increases the mean density of solids in Jupiter's feeding zone, because of an inner density wave of planetesimals produced by the rapid formation of Saturn (see Fig.~\ref{fig:CompPerfiles6NMSigma05_2}).}
\label{fig:CompDensMedias6NMSigma05}
\end{figure}

\begin{figure} 
\centering 
\includegraphics[height=0.4\textheight]{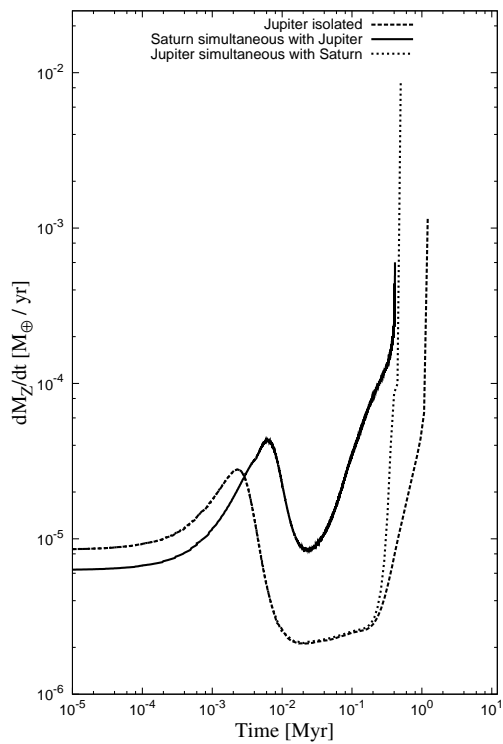} 
\caption{A comparison of the evolution of the solid accretion rate for the simultaneous formation of Jupiter and Saturn and isolated formation of Jupiter for a 6~MMSN disk with power index $p= 1/2$. We see that Saturn's rapid formation significantly increases the solid accretion rate for the simultaneous formation of Jupiter.}
\label{fig:CompTasas6NMSigma05}
\end{figure}

The rapid formation of Saturn produces a {\it density wave} of solids, responsible for the acceleration of Jupiter's formation (see Fig.~\ref{fig:CompPerfiles6NMSigma05_2}). This wave increases the mean density of solids in Jupiter's feeding zone (Fig.~\ref{fig:CompDensMedias6NMSigma05}) increasing the accretion rate of solids (Fig.~\ref{fig:CompTasas6NMSigma05}). In our model, we do not consider some potentially relevant phenomena such as planetesimals resonant trapping, planetesimals shepherding, and gap opening in the planetesimal disk (Tanaka \& Ida, \cite{ti1}, \cite{ti2}; Zhou \& Lin, \cite{zl}; Shiraishi \& Ida, \cite{si}). These phenomena could change the process of planetary formation slowing down or even inhibiting solid accretion. Unfortunately, they are very difficult to incorporate in a semi-analytical model such as the one presented here.

\section{The case of the Hayashi nebula}
\label{sec:Hayahshi_Nebula}

Finally, for the sake of completeness, we repeated the calculations for the case of the standard model of Hayashi. Since this model only differs from our own with $p=3/2$ in terms of the amount of gas, similar results may be expected because a large amount of gas implies lower relative velocities of the planetesimals, and makes the accretion of solids more efficient. The corresponding results are given below.

\subsection{Jupiter and Saturn: isolated formation}

Comparing the two considered values of the gas/solid ratio (Tables~\ref{table:IsoForSigma15} and \ref{table:IsoForHayashi}) we find the expected results. The higher the gas/solid ratio, the lower the formation timescale and the higher the core mass. The effects are clearly much larger for the case of Saturn. We now consider whether qualitatively, the main characteristics of simultaneous formation remain the same.

\subsubsection{Jupiter and Saturn: simultaneous formation}

We see (Table~\ref{table:SimForHayashi}) that, except for the case of 5~MMSN, the simultaneous formation changes quantitatively and even qualitatively relative to the results for the case of a ratio gas/solid=100 (Table~\ref{table:SimForSigma15}). For the cases of 6, 7, and 8~MMSN, the formation times of Saturn shorten, while those of Jupiter become longer and the core masses are also lowered considerably. This is because, while an increment in the amount of gas improves the efficiency of the solid accretion rates, it also accelerates the planetesimal migration.

\begin{centering}
\begin{table} 
\caption{Isolated formation of Jupiter and Saturn for the Hayashi nebula.}
\begin{tabular}{p{0.4cm}|p{0.58cm} p{0.58cm} p{0.58cm} p{0.65cm}|p{0.58cm} p{0.58cm} p{0.58cm} p{0.65cm}}
\toprule[0.8mm]

MM & & Jupiter & & & & Saturn & & \\
SN & & & & & &  & & \\ 

 \midrule[0.6mm]

 & $\Sigma_s$ 
 & $\rho_g$ 
 & $M_c$
 & $t_f$ 
 & $\Sigma_s$ 
 & $\rho_g$ 
 & $M_c$  
 & $t_f$ \\

\cmidrule[0.6mm](l){2-9} 

                5 & 12.65 & 7.51 & 30.82 & 2.58 & 5.12 & 1.43 & 17.23 & 8.77 \\
\midrule[0.3mm] 6 & 15.18 & 9.02 & 37.68 & 1.49 & 6.14 & 1.72 & 21.68 & 4.00 \\
\midrule[0.3mm] 7 & 17.71 & 10.5 & 41.67 & 0.80 & 7.17 & 2.00 & 25.30 & 2.20 \\
\midrule[0.3mm] 8 & 20.24 & 12.0 & 42.81 & 0.42 & 8.19 & 2.30 & 28.50 & 1.22 \\

\bottomrule[0.8mm] 
\end{tabular} 
\label{table:IsoForHayashi}
\end{table} 
\end{centering}

\begin{centering}
\begin{table} 
\caption{Same as Table \ref{table:IsoForHayashi} but for the simultaneous formation of Jupiter and Saturn.}
\begin{tabular}{p{0.4cm}|p{0.58cm} p{0.58cm} p{0.58cm} p{0.65cm}|p{0.58cm} p{0.58cm} p{0.58cm} p{0.65cm}}
\toprule[0.8mm]

MM & & Jupiter & & & & Saturn & & \\
SN & & & & & &  & & \\ 

 \midrule[0.6mm]

 & $\Sigma_s$ 
 & $\rho_g$ 
 & $M_c$
 & $t_f$ 
 & $\Sigma_s$ 
 & $\rho_g$ 
 & $M_c$  
 & $t_f$ \\

\cmidrule[0.6mm](l){2-9} 

                5 & 12.65 & 7.51 & 22.68 & 3.38 & 5.12 & 1.43 & 14.93 & $> 10$ \\
\midrule[0.3mm] 6 & 15.18 & 9.02 & 25.63 & 1.90 & 6.14 & 1.72 & 23.33 & 3.81 \\
\midrule[0.3mm] 7 & 17.71 & 10.5 & 29.02 & 0.86 & 7.17 & 2.00 & 27.88 & 1.95 \\
\midrule[0.3mm] 8 & 20.24 & 12.0 & 33.83 & 0.41 & 8.19 & 2.30 & 32.71 & 0.99 \\

\bottomrule[0.8mm] 

\end{tabular}
\label{table:SimForHayashi} 
\end{table} 
\end{centering}
\section{Discussion and conclusions} 
\label{sec:disc}

In the framework of the core instability hypothesis, we have considered the in situ formation of Jupiter and Saturn occurring simultaneously in a protoplanetary disk populated by planetesimals with a size distribution for which most of the mass is in small objects. We considered protoplanetary nebulae with power-law surface densities for gas and solids, and assumed that the gas component of the disk dissipates following an exponential law with a characteristic timescale of 6~Myr. 

\begin{figure} 
\centering 
\includegraphics[height=0.4\textheight]{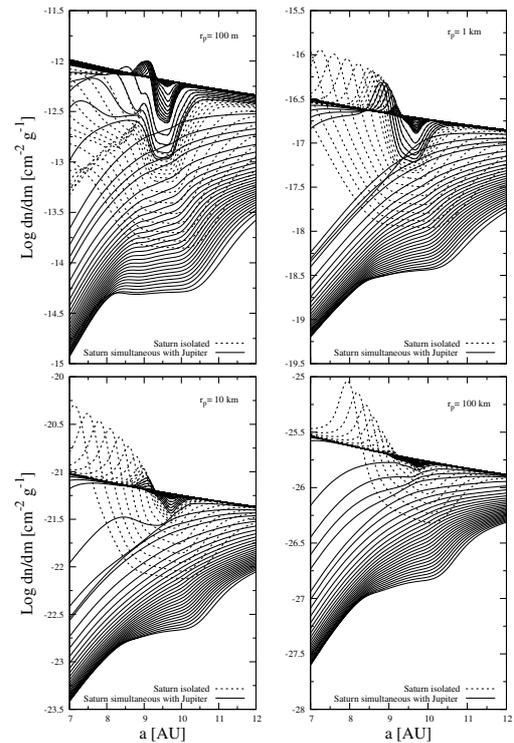} 
\caption{Comparison of the time evolution of planetesimal density profiles for a 8~MMSN disk with power index $p= 3/2$ in Saturn's neighborhood for its isolated formation, and also its simultaneous formation with Jupiter. As time advances, the density profiles decrease in both cases; however, profiles evolution are very different. Planetesimals of 1, 10, 100~km in size produce a density wave for Saturn's isolated formation. This density wave is inhibited when Saturn forms simultaneously with Jupiter: for all sizes of planetesimals, the evolution of density profiles decreases more quickly when Saturn forms in the presence of Jupiter.} 
\label{fig:CompPerfiles8NM} 
\end{figure}

In the first instance, we calculated the isolated and simultaneous formation of Jupiter and Saturn for a standard nebula with profile $\Sigma \propto r^{-3/2}$ but with a lower gas/solid ratio than that adopted by Hayashi~(\cite{hayashi}), following the work by Mordasini et al.~(\cite{mordasini}). We quantitatively analyzed how the isolated formation of Jupiter (Saturn) is affected when it occurs simultaneously with Saturn (Jupiter). For the isolated formation, we have found that Jupiter and Saturn achieved its final masses on a timescale in agreement with observational estimate for 6 to 10~MMSN (the corresponding disk masses are between 0.003~$M_{\odot}$ and 0.05~$M_{\odot}$). However, the final core masses are higher than estimates in most of the cases (Table~\ref{table:IsoForSigma15}). The most important result is that Jupiter's formation timescales are somewhat shorter than the corresponding ones for Saturn. We have found that the rapid formation of Jupiter inhibits -or largely increases- the timescale of Saturn's formation when they grow simultaneously inside the disk (Table~\ref{table:SimForSigma15}). Jupiter's formation increases the migration mean velocities of the planetesimals at Saturn's feeding zone (Figs.~\ref{fig:CompVelos6NM} and \ref{fig:CompVelos8NM}). This phenomena causes the timescale of solids accretion rate to becomes longer than the planetesimal migration timescale such that, Saturn's solid accretion rate becomes less efficient (Figs.~\ref{fig:CompTasas6NM} and \ref{fig:CompTasas8NM}). In Fig.~\ref{fig:CompPerfiles8NM}, we show that the presence of Jupiter significantly modifies the time evolution of solids in Saturn's neighborhood and, in turn, modifies Saturn formation.

We then explored smoother profiles in trying to solve the problem of the different formation timescales. Accretion $\alpha$~disks predict profiles of $\Sigma \propto r^{-1}$. We used this profile and another proposed by Lissauer~(\cite{lissauer87}) -$\Sigma \propto r^{-1/2}$- that could solve formation timescales of the outer Solar System giant planets. We normalized these profiles at Jupiter's position (5.2~AU) so that, at this position, the surface densities of solids and gas, and the volumetric density of gas in the mid plane disk are the same for all considered profiles ($p=3/2,1,1/2$). The smoother the profiles, the higher the density of solids at Saturn's position, and the closer the similarity in the formation timescales.

For the profile $\Sigma \propto r^{-1}$, we found that the formation timescales of Jupiter and Saturn are more similar to those for isolated formation. In this case, the simultaneous formation of Jupiter and Saturn is practically does not unchanged with respect to the corresponding isolated formation of each planet. However, for the profile $\Sigma \propto r^{-1/2}$ the temporal order of the phenomena is inverted. In most cases, Saturn forms before Jupiter. The rapid formation of Saturn induces a density wave of solids (Fig.~\ref{fig:CompPerfiles6NMSigma05_2}) that significantly favors Jupiter's formation (Fig.~\ref{fig:CompMasas6NMSigma05}) . This wave increases the amount of solids in Jupiter's feeding zone (Fig.\ref{fig:CompDensMedias6NMSigma05}), such that the solid accretion rate becomes more efficient (Fig.\ref{fig:CompTasas6NMSigma05}).

\begin{figure} 
\centering 
\includegraphics[height=0.5\textheight]{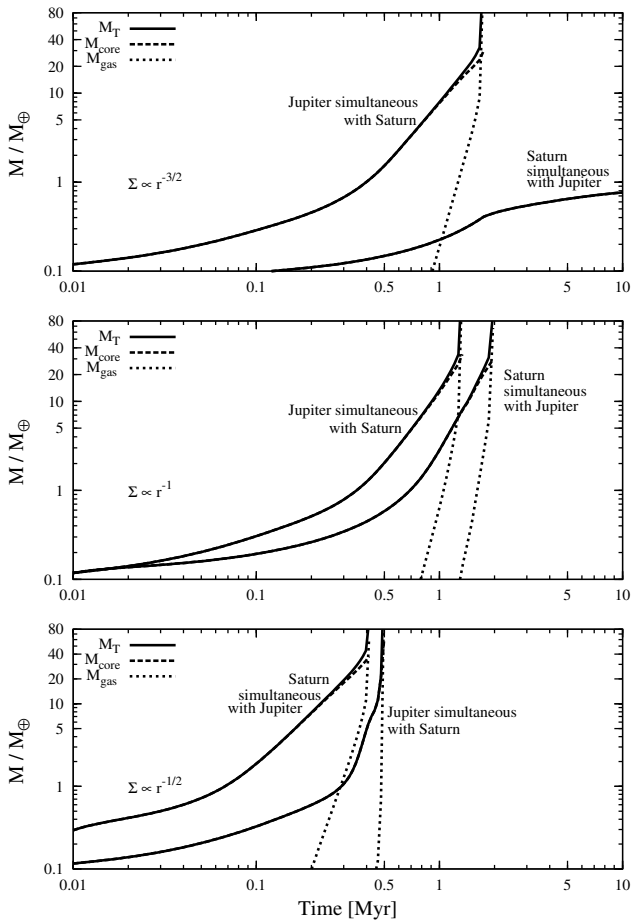}
\caption{Comparison of the simultaneous formation of Jupiter and Saturn for a 6~MMSN disk with different power indices ($p=~3/2;1;1/2$). We see that simultaneous formation of Jupiter and Saturn changes as the power index decreases.}
\label{fig:CompMasasPerfiles6NM}
\end{figure}

In Fig.~\ref{fig:CompMasasPerfiles6NM}, we compare the simultaneous formations of Jupiter and Saturn for the case of a 6~MMSN disk with different power indices ($p= 3/2;1;1/2$). For the case of $p=3/2$, Jupiter inhibits Saturn's formation. For smoother profiles ($p= 1;1/2$), the process becomes more efficient, for various reasons. First, at Saturn's position there is a higher density of solids and gas. In addition, as power indices decrease the mass of solids grows outward, and due to planetesimal migration the replenishment of solids in the feedings zones is more efficient.

Finally, we repeated the calculations for the standard model of Hayashi. Because Hayashi's nebula is more massive than that of our model with $p= 3/2$ -only differing in the amount of gas- our results were expected to be qualitatively similar. This was found to be true for the isolated formation of Jupiter and Saturn (Table \ref{table:IsoForHayashi}), where formation timescales decrease and final core masses increase. This is because a greater amount of gas leads to the accretion rate of solids becoming more efficient. However, the properties of the simultaneous formation of Jupiter and Saturn differ qualitatively and quantitatively in the case of our model with $p= 3/2$. While for our model Jupiter inhibits Saturn's formation, for the Hayashi's nebula the formation times of Saturn are decreased by the presence of Jupiter. In addition, Jupiter final core masses are significantly decreased by the presence of Saturn.

\begin{figure} 
\centering 
\includegraphics[height=0.5\textheight]{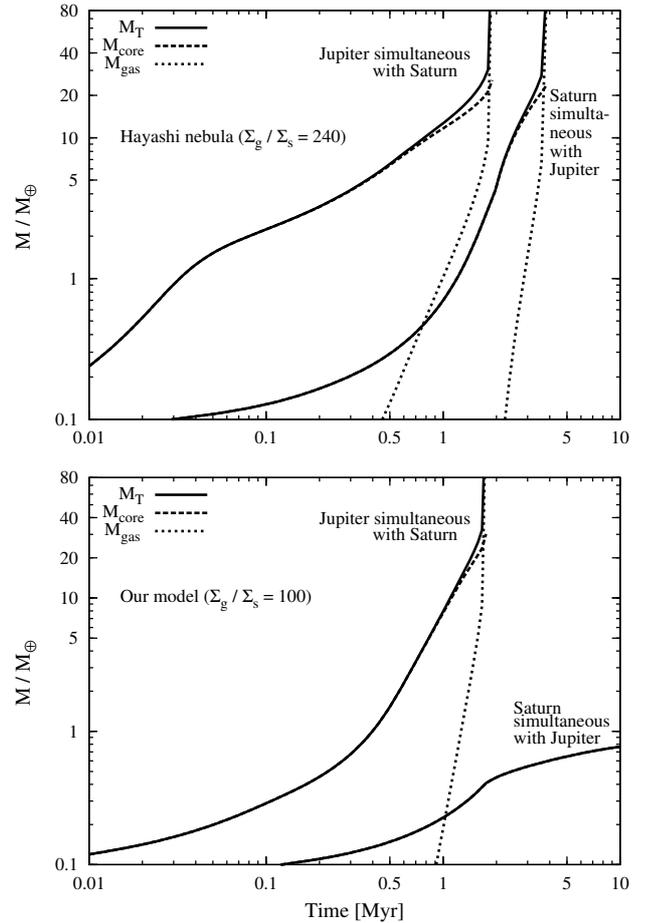}
\caption{Comparison of the simultaneous formation of Jupiter and Saturn for a 6~MMSN disk for the Hayashi nebula ($\Sigma_{g}/\Sigma_{s}= 240$) and our model ($\Sigma_{g}/\Sigma_{s}= 100$). The formation processes are very different in each case.} 
 \label{fig:CompMasasHayashi6NM} 
\end{figure}

In Fig.~\ref{fig:CompMasasHayashi6NM}, we show the differences between the simultaneous formations of Jupiter and Saturn for a 6~MMSN disk between the Hayashi nebula and our model with $p= 3/2$. The gas/solid ratio clearly has an important role in the formation process. Different gas/solid ratios may change qualitatively and quantitatively the whole formation of a planetary system.

The primary hypothesis of our model is that most of the planetesimal mass is in small objects. Most of the works that simulate giant planet formation do not consider oligarchic growth regime. In these works, a much faster time-dependent regime is adopted for the growth of the core (Pollack et al., \cite{pollack}; Alibert et al., 2005 \cite{alibert a}, \cite{alibert b}; Hubickyj et al., \cite{hubickyj}; Dodson-Robinson et al. \cite{dr}). With this solid accretion rate, the final core mass of the giant planet is quickly achieved. Relative velocities are slower and the solid accretion rate becomes more efficient, so planetesimals of 100~km in size are usually used. While some studies predict this type of planetesimals (Johansen et al., \cite{johansen}; Morbidelli et al., \cite{morbidelli}), N-body simulations predict that when the embryos reach a mass similar to the Moon, the runaway regime switches to oligarchic regime (Ida \& Makino, \cite{ida}; Kokubo \& Ida, \cite{kokubo1}; \cite{ki2000}; \cite{kokubo3}). We note that if our primary hypothesis is relaxed, and most of the planetesimal mass does not reside in small objects -or larger planetesimals are considered- formation timescales become longer than disk lifetimes.
 

\begin{acknowledgements}
We are especially grateful to A. Fortier for the close collaboration and the lively discussions.	
\end{acknowledgements}


\end{document}